\documentclass{article} % For LaTeX2e
\usepackage{iclr2023_conference,times}

\usepackage{hyperref}
\usepackage{url}
\usepackage{graphicx}
\usepackage{booktabs}
\usepackage{multirow}
\usepackage{array}
\usepackage{bbding}
\usepackage{subcaption}
\usepackage{algorithm}
\usepackage{algorithmic}
\usepackage{wrapfig}
\usepackage{colortbl}
\usepackage{amsmath,amsfonts,bm}

\definecolor{mydarkblue}{rgb}{0,0.08,0.45}
\definecolor{mydarkgreen}{RGB}{0, 139, 69}
\definecolor{mygreen2}{RGB}{0 205 0}
\definecolor{mybrown}{RGB}{139 69 19}
\definecolor{mybrown2}{RGB}{128 70 27}
\definecolor{mybrown3}{RGB}{107 68 35}
\definecolor{MAEblue}{RGB}{47 112 182}

\hypersetup{
	colorlinks=true,
	urlcolor=magenta,
	citecolor=MAEblue,
}
\definecolor{mycyan}{cmyk}{.3,0,0,0}

\def\myname{SDMuse}

\title{\myname{}: Stochastic Differential Music Editing and Generation via Hybrid Representation}

\author{Chen Zhang\thanks{This work was done during an internship at Sea AI Lab, Singapore.}~~\textsuperscript{\textnormal{1}~\textnormal{2}}, Yi Ren~\textsuperscript{\textnormal{2}}, Kejun Zhang~\textsuperscript{\textnormal{1}}, Shuicheng Yan~\textsuperscript{\textnormal{2}}, \\
\textsuperscript{1}Zhejiang University, China \\
\textsuperscript{2}Sea AI Lab, Singapore\\
}

\iclrfinalcopy % Uncomment for camera-ready version, but NOT for submission.
\begin{document}

\maketitle

\begin{abstract}
% Song-level symbolic music generation with high quality and controllability is one of the most important but challenging tasks in automatic music composition. Previous works of song-level music generation tackles long-term structure modeling mainly based on the representation of MIDI event sequence and auto-regressive generation models, which may suffer from error accumulation and restricted by the very long token sequence. On the other hand, existing works about controllable music generation tend to use some explicit control signals which need to be annotated in datasets, thus are still limited in the ways of controlling and have problems in collaborating with humans flexibly. In this work, to address aforementioned problems, we propose \myname{}, a stochastic differential music editing and generation method, which consists of two stages: 1) stochastic differential editing and generation of music score; and 2) auto-regressive generation and refinement on music performance to ensure the high quality and different-level controllability at the same time. \myname{} uses hybrid representation, thus can take advantages of two symbolic music representations (MIDI event token and piano-roll). We conduct experiments on \textit{ailabs1k7} pop music dataset and evaluate the output music both objectively and subjectively in terms of \textit{quality} and \textit{controllability}, and the experimental results demonstrate the effectiveness of \myname{}. The generated samples can be found at our demo page \url{https://\myname.github.io}.

While deep generative models have empowered music generation, it remains a challenging and under-explored problem to edit an existing musical piece at fine granularity. In this paper, we propose \myname{}, a unified \underline{S}tochastic \underline{D}ifferential \underline{Mus}ic \underline{e}diting and generation framework, which can not only compose a whole musical piece from scratch, but also modify existing musical pieces in many ways, such as combination, continuation, inpainting, and style transferring. The proposed \myname{} follows a two-stage pipeline to achieve music generation and editing on top of a hybrid representation including pianoroll and MIDI-event. In particular, \myname{}~first generates/edits pianoroll by iteratively denoising through a stochastic differential equation (SDE) based on a diffusion model generative prior, and then refines the generated pianoroll and predicts MIDI-event tokens auto-regressively.
% \myname{} is based on a diffusion model generative prior, synthesizing a musical piece by iteratively denoising through a stochastic differential equation. 
%quanhong: synthesize
%Considering the pianoroll and MIDI-event are suitable for editing and modeling respectively, 
% We design a two-stage pipeline, including pianoroll and MIDI-event generation, which takes advantage of two kinds of symbolic music representations (we call it hybrid representation).
We evaluate the generated music of our method on \textit{ailabs1k7} pop music dataset in terms of quality and controllability on various music editing and generation tasks. Experimental results demonstrate the effectiveness of our proposed stochastic differential music editing and generation process, as well as the hybrid representations. 

% PURE TEXT VERSION
% While deep generative models have empowered music generation, it remains a challenging and under-explored problem to edit an existing musical piece at fine granularity. In this paper, we propose SDMuse, a unified stochastic differential music editing and generation framework, which can not only compose a whole musical piece from scratch, but also modify existing musical pieces in many ways, such as combination, continuation, inpainting, and style transferring. The proposed SDMuse follows a two-stage pipeline to achieve music generation and editing on top of a hybrid representation including pianoroll and MIDI-event. In particular, SDMuse first generates/edits pianoroll by iteratively denoising through a stochastic differential equation (SDE) based on a diffusion model generative prior, and then refines the generated pianoroll and predicts MIDI-event tokens auto-regressively. We evaluate the generated music of our method on ailabs1k7 pop music dataset in terms of quality and controllability on various music editing and generation tasks. Experimental results demonstrate the effectiveness of our proposed stochastic differential music editing and generation process, as well as the hybrid representations. 
\end{abstract}

\section{Introduction}
\label{sec:intro}
With the development of deep learning and generative models,
%~\citep{kingma2013auto,goodfellow2020generative,nichol2021glide}.
automatic music composition has received much research attention~\citep{dong2018musegan,huang2020pop,ren2020popmag,ju2021telemelody,wu2021musemorphose,zhang2022relyme}, and also has a lot of successful applications, such as movie soundtracks, virtual singers, and auxiliary composing. 
However, as the aesthetics of music are diverse for different groups of people, or even for each individual, there is no musical piece in the world that simultaneously satisfies everyone's taste for any scenario.
% music composition task in general does not possess any single well-defined solution.
In many cases, one may feel unsatisfied with certain bars of a musical piece, if not with the whole piece, and he or she would like to edit them, which is impossible except for a professional such as a music composer. 
AI solutions like symbolic music generation methods~\citep{huang2020pop,hsiao2021compound} can help, but they would regenerate a completely new piece that may be totally different from the former one without preserving those ``satisfying" parts. 
The same thing happens when someone wants to extend an existing musical piece, modify the style or combine several pieces.
Therefore, besides the effort on improving generative quality~\citep{ren2020popmag,ju2021telemelody} in a broad sense, it is also crucial and much demanded to study how to edit and modify existing musical pieces.
%in the field of automatic music composition.
% Aesthetics of music are highly personalized, and hence unlike other machine learning tasks (e.g. machine translation, image classification), the music composition task does not have a consistent standard for the best results.
% Given the same musical piece, different people may have different evaluations related to their own experiences and personalities, so it is almost impossible for music generation models to generate musical pieces that satisfy everyone at once. 
% Also, when music generation is used in industry, a single model cannot directly generate suitable musical compositions that meet various commercial demands.
% Therefore, in addition to focusing on the generative \textit{quality}~\citep{huang2020pop,hsiao2021compound} in a broad sense, the study of how to edit and adjust existing musical pieces is also a very important part in the field of automatic music composition.
% Despite tremendous progress in deep generative models, which have enhanced different kinds of automatic music composition tasks, editing existing musical pieces (especially at a fine-grained level) is still quite challenging and under-studied. 

There are only a few works studying music editing tasks, mainly focusing on global music style transfer~\citep{cifka2020groove2groove,wu2021musemorphose} or music genre transfer~\citep{brunner2018symbolic}. The most related work \citep{wu2021musemorphose} to ours achieves music style transfer with Transformer VAE on pre-defined musical attributes like rhythmic intensity and polyphony. The musical attributes are song-level, prohibiting users from editing certain part of a musical piece. In addition, users can only control and edit the pre-defined attributes, which severely limits its application. 
Other works~\citep{cifka2020groove2groove,brunner2018symbolic} have similar limitations. 
So far, no one has explored editing musical pieces at fine granularity in the same flexible way as image editing,
%~\citep{jo2019sc,dekel2018sparse}, 
such as inpainting~\citep{pathak2016context} and outpainting~\citep{wang2014biggerpicture}.

\begin{figure}[!t]
	\centering
	\begin{subfigure}[b]{0.3\textwidth}
	    \captionsetup{justification=centering}
		\centering
	    \includegraphics[width=\textwidth,trim={0cm 0cm 0cm 0cm}, clip=true]{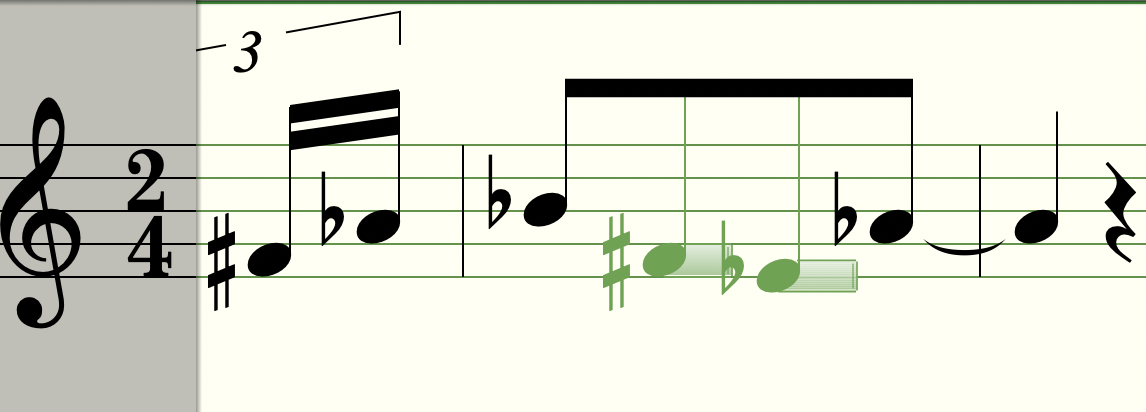}
	    \vspace{0.1cm}
	    \caption{Music sheet.}
	    \label{fig:music_sheet}
	\end{subfigure}
	\hspace{0.1cm}
	\begin{subfigure}[b]{0.34\textwidth}
	    \captionsetup{justification=centering}
		\centering
	    \includegraphics[width=\textwidth,trim={0cm 0cm 0cm 0cm}, clip=true]{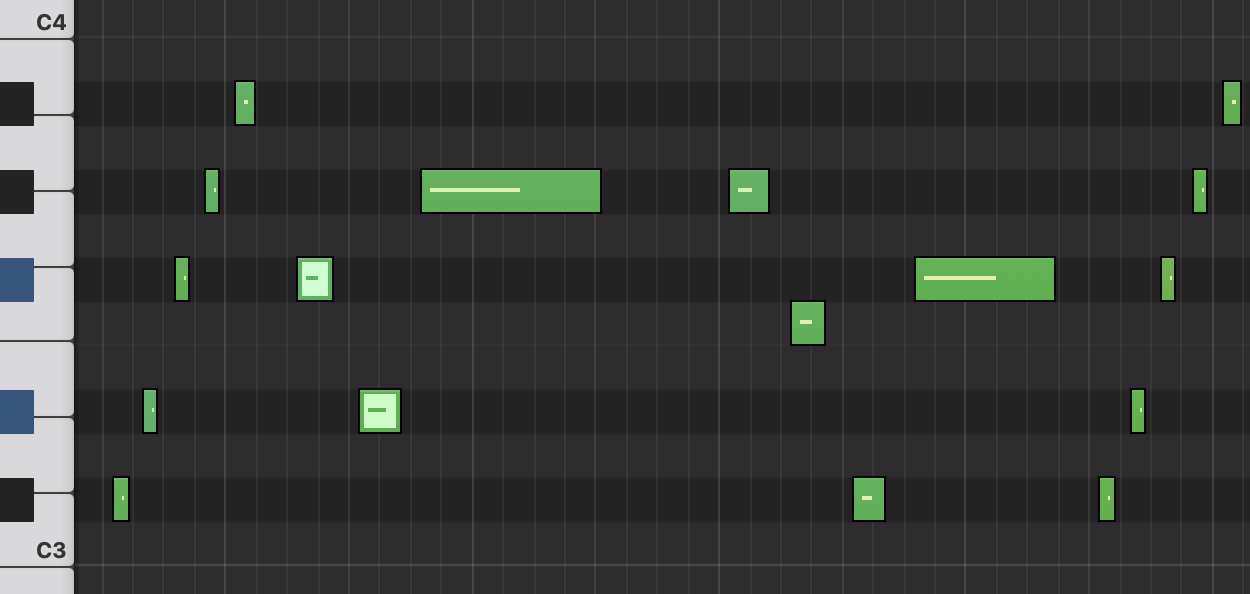}
	   % \vspace{0.1cm}
	    \caption{Pianoroll.}
	    \label{fig:pianoroll}
	\end{subfigure}
	\hspace{0.1cm}
	\begin{subfigure}[b]{0.26\textwidth}
	    \captionsetup{justification=centering}
		\centering
	    \includegraphics[width=\textwidth,trim={0cm 0cm 0cm 0cm}, clip=true]{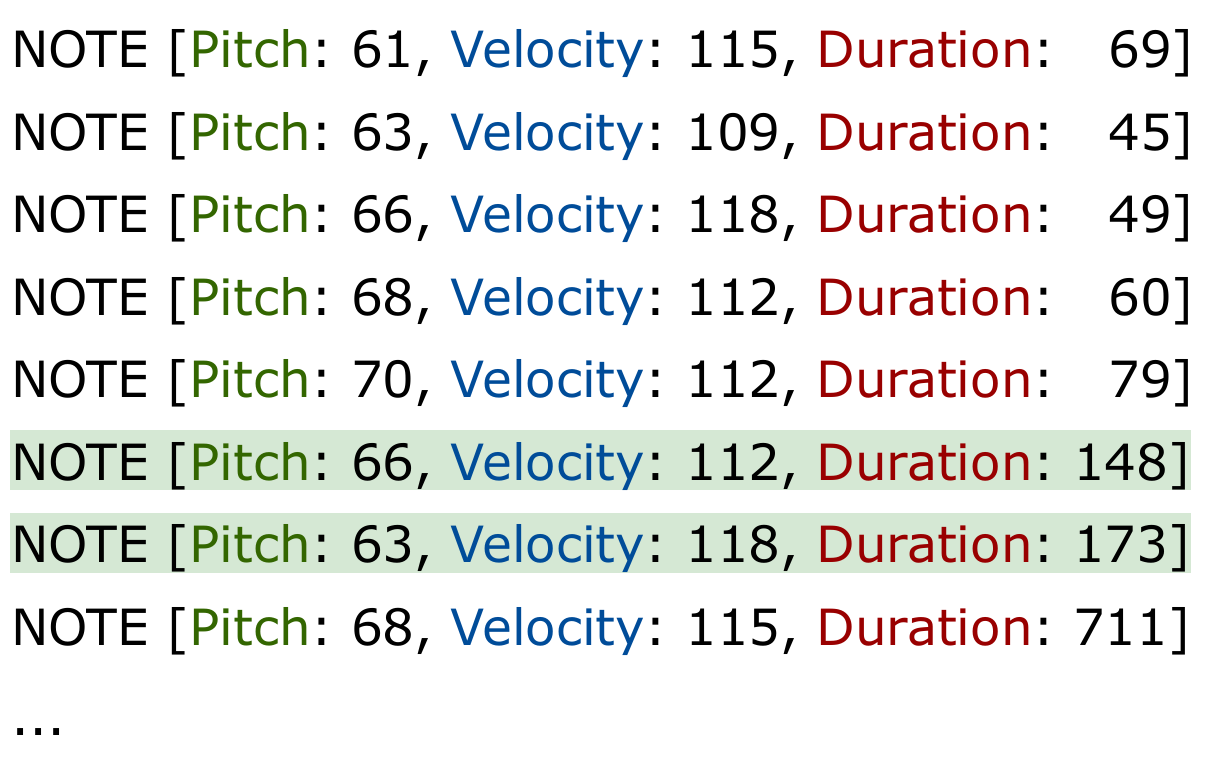}
	    \caption{MIDI-event tokens.}
	    \label{fig:midi_event}
	\end{subfigure}
	\caption{The illustrations of different symbolic music representations.}
\end{figure}

Existing state-of-the-art music generation works~\citep{hsiao2021compound,ren2020popmag,zhang2022relyme} with high generative quality are based on MIDI-event representation (an illustration of MIDI-event is in Figure~\ref{fig:midi_event}).
As validated by the success of these methods, the MIDI-event representation is appropriate for the model to generate and model music performance details such as velocity, but not for humans to perceive and edit since the data format is not intuitive enough. Another widely used music representation pianoroll (an illustration of pianoroll is in Figure~\ref{fig:pianoroll}) is closer to the way of human perception, but the methods~\citep{dong2018musegan,yang2017midinet} based on it cannot achieve comparable generation quality with MIDI-event-based ones. 
In this work, we consider using both of the representations, which we call hybrid representation, to achieve fine-grained music editing.

Specifically, we define a series of fine-grained music editing tasks and propose a unified \underline{S}tochastic \underline{D}ifferential \underline{Mus}ic \underline{e}diting and generation framework via hybrid representation, named \myname{}.
\myname{} can not only generate a whole musical piece from scratch either unconditionally or conditioned on some control signals (such as chord progression), but also edit existing musical pieces in different ways, including stroke-based generation/editing, inpainting, outpainting, combination and style transferring. 
As pianoroll is easier for humans to understand and edit, and MIDI-event sequence is more suitable for generation, we design a two-stage pipeline based on a hybrid representation.
In the first stage, a diffusion model generative prior is applied and the pianorolls are obtained by iteratively denoising through a stochastic differential equation (SDE). The progressive generation feature of our diffusion model allows us to implement a series of music editing operations in this stage.
In the second stage, the generated pianoroll will be refined with more precise music performance details (velocity, fine-grained onset position, etc.) by generating the final MIDI-event sequence in an auto-regressive manner, enjoying the benefits of MIDI-event representation for high-quality music generation.

We evaluate \myname{} on \textit{ailabs1k7} pop music dataset~\citep{hsiao2021compound} in terms of \textit{quality} and \textit{controllability} in various music editing and generation tasks.
Both objective and subjective results demonstrate the effectiveness of the stochastic differential process and hybrid representation in \myname{}. To our best knowledge, we are the first one to formulate and address fine-grained music editing, which aims to edit existing musical pieces at fine granularity according to diverse demands and provide humans with new ways to collaborate with models during music composition. The generated samples can be found on our demo page \url{https://\myname.github.io/posts/sdmuse/}.

\section{Background}
\label{sec:background}

\subsection{Symbolic music representation}
\label{sec:bg_music_repr}
Most previous symbolic music generation works are based on two most common music representations: pianoroll and MIDI-event. Pianoroll-based approaches~\citep{yang2017midinet,dong2018musegan} use pianorolls to represent music scores, with the horizontal axis representing time and the vertical axis representing pitch. Pianoroll is just like an image, so pianoroll-based methods use image-based operations to generate music. Pianoroll is closer to the way of humans perceiving musical pieces, making it more suitable for being understood and edited by humans.
However, most state-of-the-art music generation works are MIDI-event-based approaches~\citep{huang2020pop,hsiao2021compound}. They convert a musical piece to a MIDI-event token sequence, and use methods from natural language processing to deal with the token sequence. MIDI-event-based methods can better learn the temporal dependency between different musical events, so as to show more robust generation performance.
We list the advantages and disadvantages of these two representations as shown in Table~\ref{tab:music_repr} and explain them in detail in Appendix~\ref{app:music_repr}. To summarize, pianoroll is more appropriate for extracting and controlling perceptive information like structure, while the MIDI-event sequence is more ideal for generating and modeling precise music performance details, such as velocity and fine-grained onset position.
In this work, we propose \myname{} with a two-stage pipeline, including pianoroll and MIDI-event generation stages that apply hybrid representation to take advantage of these two symbolic music representations.

\begin{table}[t]
    \small
	\centering
	\vspace{-0.5cm}
	\caption{Differences between MIDI-event and pianoroll.}
	\begin{tabular}{m{2cm} | l l}
		\toprule
		  & \multicolumn{1}{c}{MIDI-event} & \multicolumn{1}{c}{Pianoroll} \\
		\midrule
		 & \Checkmark ~precise details (velocity, precise position) & \Checkmark ~prior music information \\
		\multirow{2}{*}{Pros \& Cons} & \Checkmark ~regard a note as the unit & \Checkmark ~reflect music structure directly \\
		 & \Checkmark ~robust to generate music pieces & \Checkmark ~easy to control and adjust \\
         & \XSolidBrush ~need more data to learn the embedding & \XSolidBrush ~treat onset and other pos samely\\
        \midrule
		Suitable Usage Scenarios & generating \& modeling & perceiving \& editing \\
        \bottomrule
	\end{tabular}
	\label{tab:music_repr} 
\end{table}
\subsection{Symbolic music editing}
\label{sec:bg_music_edit_gen}
Though tremendous progress is made in symbolic music generation and other automatic music composition tasks, there are only a handful of works regarding symbolic music editing tasks. 
The existing music editing works mainly focus on global music style transfer~\citep{cifka2020groove2groove,wu2021musemorphose} and music genre transfer~\citep{brunner2018symbolic}. \citet{cifka2020groove2groove} proposed a system for music accompaniment style transfer, generating accompaniment with the content from content input and the style from style input. 
\citet{wu2021musemorphose} changed the style of a musical piece by given song-level musical attributes (e.g. rhythmic intensity and polyphony).
\citet{brunner2018symbolic} applied GAN-based methods from the field of computer vision to transfer a musical piece from source genre to target genre.
However, these works can only edit the music at the song level with pre-defined attributes or labels, limiting the ways of interaction. This work aims to implement fine-grained music editing and achieve flexible collaboration between humans and models during music composition.

\subsection{Stochastic differential equations (SDE) for editing}
\label{sec:bg_sde}
To recover the data from noise, \citet{song2020score} proposed a stochastic differential equation (SDE) to smoothly transform a complex data distribution to a known prior distribution by slowly injecting noise.
Similar to diffusion probabilistic models~\citep{ho2020denoising,liu2021diffsinger,mittal2021symbolic}, SDE-based generative models can be used to convert an initial Gaussian noise vector to a data point in real-world data distribution.
As described in \citet{song2020score}, we denote $\mathbf{x}(t) \in \mathbb{R}^d$, where $t \in [0,T]$ represents time. Suppose that $\mathbf{x}(0) \sim p_{data}$ is a sample from data distribution, and $\mathbf{x}(T) \sim p_{T}$ is from prior distribution. The forward SDE process can be formulated as:
\begin{equation}
    \mathrm{d}\mathbf{x}(t) = \mathbf{f}(\mathbf{x}, t)\mathrm{d}t + g(t)\mathrm{d}\mathbf{w},
    \label{eq:forward_sde}
\end{equation}
where $\mathbf{w}$ is a standard Brownian motion, and $\mathbf{f}(\cdot,t)$, $g(\cdot)$ are the drift coefficient and the diffusion coefficient of $\mathbf{x}(t)$ respectively.
And the reverse SDE~\citep{anderson1982reverse} is:
\begin{equation}
    \mathrm{d}\mathbf{x}(t) = [\mathbf{f}(\mathbf{x}, t) - g(t)^2 \nabla_{\mathbf{x}}\mathrm{log}p_t(\mathbf{x})] \mathrm{d} t + g(t) \mathrm{d} \overline{\mathbf{w}},
    \label{eq:reverse_sde}
\end{equation}
where $\overline{\mathbf{w}}$ is a standard Wiener process and $\nabla_{\mathbf{x}}\mathrm{log}p_t(\mathbf{x})$ is the noise-perturbed score function.
% According to~\citet{song2020score}, we denote $\mathbf{x}(t) \in \mathbb{R}^d$, where $t \in [0,T]$ represents time. Suppose that $\mathbf{x}(0) \sim p_{data}$ is a sample from data distribution, $\mathbf{x}(T) \sim p_{T}$ is from prior distribution. The forward VP-SDE process can be formulated 
There are two different SDEs according to different noise perturbations: Variance Exploding SDE (VE-SDE) and Variance Preserving SDE (VP-SDE). In this paper, we use VP-SDE for conducting experiments to verify our framework.
The forward process of VP-SDE is:
\begin{equation}
    \mathrm{d}\mathbf{x}(t) = -\frac{1}{2}\beta(t)\mathbf{x}(t)\mathrm{d}t + \sqrt{\beta(t)} \mathrm{d}\mathbf{w}(t),
    \label{eq:forward_vpsde}
\end{equation}
where $\beta(t)$ is a positive noise function. Denote the learned score model as $\bm{s_\theta}(\mathbf{x}(t), t))$. The reverse VP-SDE process can be solved by following the iteration rule:
\begin{equation}
    \mathbf{x}_{n-1} = \frac{1}{\sqrt{1-\beta(t_n) \Delta t}} (\mathbf{x}_n + \beta(t_n) \Delta t \bm{s_\theta}(\mathbf{x}(t_n), t_n)) + \sqrt{\beta(t_n) \Delta t} ~\mathbf{z_n},
    \label{eq:reverse_vpsde}
\end{equation}
where $\mathbf{x}_N, \mathbf{z}_n \sim \mathcal{N}(\mathbf{0}, \bm{\mathit{I}})$, $n = N, N-1, \cdots, 1$, and $\Delta t$ is time interval between $\mathbf{x}_n$ and $\mathbf{x}_{n-1}$.

In order to synthesize and edit images, \citet{meng2021sdedit} “hijacked” the generative process of SDE-based generative models. Specifically, they added noise to smooth the given stroke paintings or images with stroke-edit to smooth out distortions but preserving the overall image structure. Then they used the noisy input to initialize the SDE and progressively remove the noise. Inspired by \citet{meng2021sdedit}, we build our \myname{} to edit and generate musical pieces through VP-SDE.

\section{Methods}
\label{sec:methods}
In this section, we first give a pipeline overview of \myname{} and then describe the design details of the pianoroll and MIDI-event generation stages respectively. Finally, we introduce the formulation and process of various fine-grained music editing tasks.

\subsection{Pipeline overview}
\begin{figure}
\vspace{-0.5cm}
    \centering
    \begin{subfigure}[t]{0.32\textwidth}
	    \captionsetup{justification=centering}
		\centering
	    \includegraphics[width=\textwidth,trim={0cm 0cm 0cm 0cm}, clip=true]{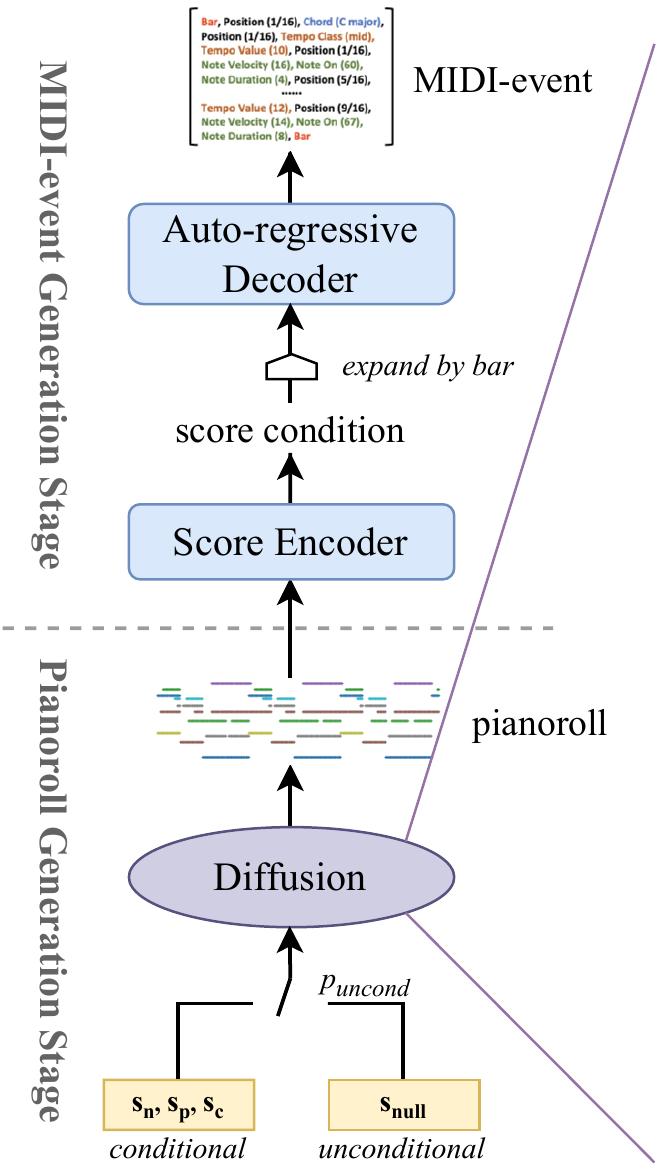}
	    \caption{\myname{} pipeline.}
	    \label{fig:pipeline}
	\end{subfigure}
	\hspace{-8pt}
	\begin{subfigure}[t]{0.66\textwidth}
	    \captionsetup{justification=centering}
		\centering
	    \includegraphics[height=0.83\textwidth,trim={0cm 0cm 0cm 0cm}, clip=true]{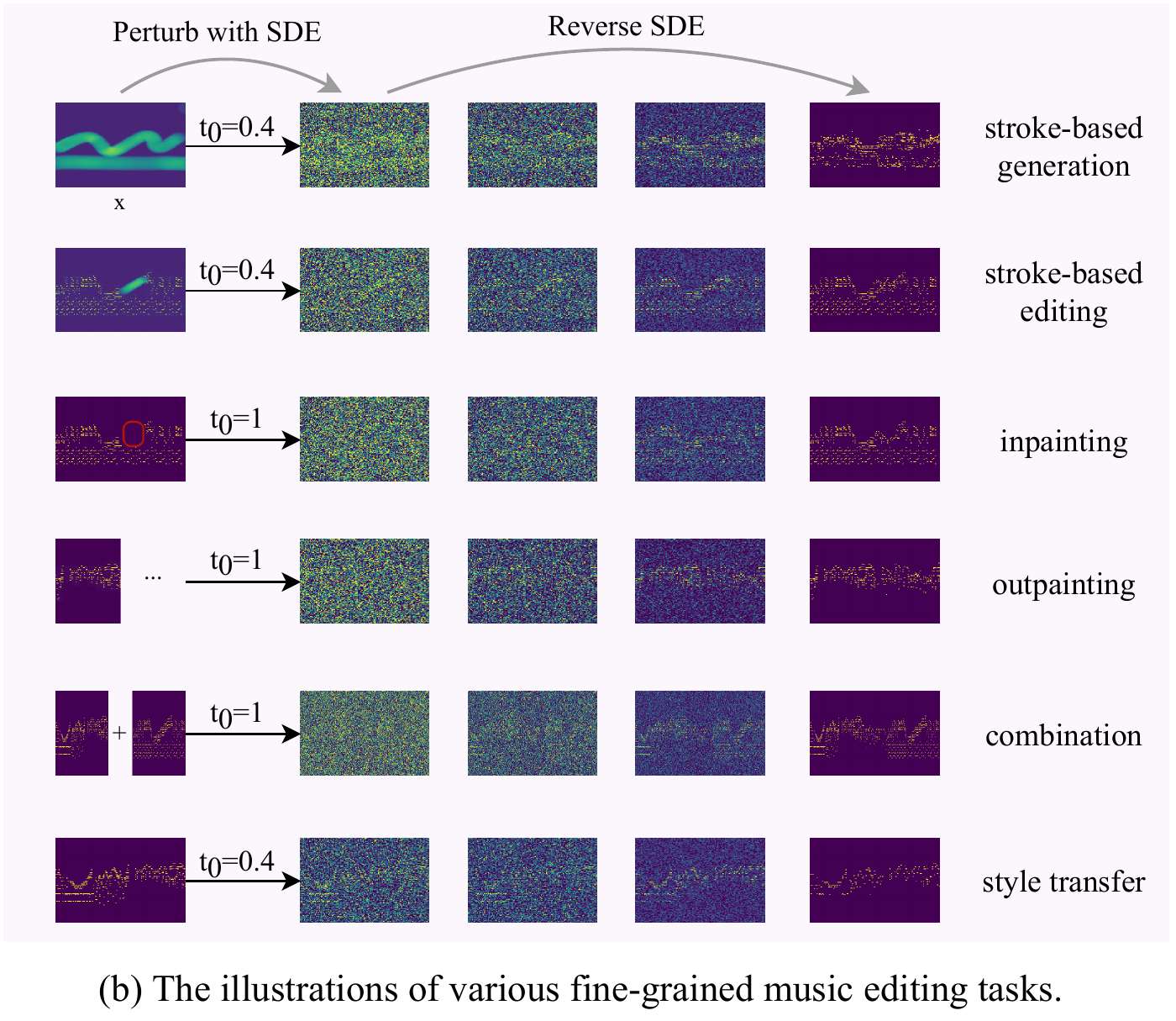}
	    \caption{The illustrations of various fine-grained music editing tasks.}
	    \label{fig:editing_tasks}
	\end{subfigure}
    \caption{Left part is the \myname{} pipeline containing two generation stages. Right part shows the process of various fine-grained music editing tasks. Best viewed in color mode with zoom-in.}
    \label{fig:main_fig}
\end{figure}

The overall pipeline of \myname{} is shown in Figure~\ref{fig:pipeline}, which consists of two consecutive stages: 1) pianoroll generation stage, which is based on a conditioned diffusion model generative prior and synthesizes pianorolls from scratch or edits existing pianorolls by iteratively denoising through SDE~\citep{song2020score}; 2) MIDI-event generation stage, which generates MIDI-event sequences from the output pianorolls of the first stage by refining them with precise music performance details auto-regressively. These two stages can be trained separately and all condition signals in the first stage are at fine granularity, which can be extracted from the musical piece itself without extra data annotation process. The diffusion probabilistic model in the pianoroll generation stage enables \myname{} not only to compose the whole musical pieces from scratch unconditionally or conditioned on given control signals (e.g. note density), but also to modify existing musical pieces in different ways. 
The details of these two stages are introduced in the following subsections and some details of the model architectures are described in Appendix~\ref{app:model_arch_details}.
% \begin{figure}
%     \centering
%     \includegraphics[width=\textwidth,trim={0cm 0cm 0cm 0cm}, clip=true]{figs/pipeline.pdf}
%     \caption{The left part is the \myname{} pipeline, which contains two generation stages. The right part shows the process of various fine-grained music editing tasks. Best viewed in color mode and with zoom-in.}
%     \label{fig:pipeline}
% \end{figure}

\subsection{Pianoroll generation stage}
\label{sec:score_gen}
% The pianoroll generation stage aims to generate or edit music scores in pianoroll format unconditionally or conditioned on the given control signals. This stage is based on a diffusion probabilistic model and synthesizes pianorolls by progressively denoising through SDE~\citep{song2020score}. 

\subsubsection{Training of diffusion probabilistic model}
As shown in Figure~\ref{fig:main_fig}, we involve several fine-grained control signals: note density $\bm{c}_n$, pitch distribution $\bm{c}_p$, and chord progression sequence $\bm{c}_c$ during the training process of the diffusion model to enable unconditional and conditional music generation/editing at the same time. 
These control signals can be extracted from the musical piece itself, and the way of extraction is listed in Appendix~\ref{app:control_signals}.
Given these control signals $\bm{c}_n$, $\bm{c}_p$, and $\bm{c}_c$, we can train a conditional diffusion model with the pairs of pianorolls and the corresponding control signals.
In order to integrate unconditional and conditional music generation into the same model without an extra training process, we introduce a combined training strategy which can switch freely between two generation settings inspired by~\citet{nichol2021glide}. 
Specifically, as illustrated in Figure~\ref{fig:pipeline}, we use the conditional training paradigm, but assign all control signals to a specific out-domain value ($\bm{c}_{null}$) with a certain probability $p_{uncond}$ to train the model for unconditional music generation scenario. 
As mentioned in Section~\ref{sec:bg_music_repr}, one of the drawbacks of pianoroll representation is indiscriminate treatment of note onsets and other positions, which does not match realistic scenarios and affects the robustness of generation.
To tackle this problem, we convert the pianoroll to onsetroll by only keeping onset information as described in Appendix~\ref{app:onsetroll} for the training process of the diffusion model.

\subsubsection{Generation from scratch}
Starting from samples of $\mathbf{x}(T) \sim p_T$, where $p_T$ is the prior distribution, we can generate musical pieces of $\mathbf{x}(0) \sim p_{data}$ unconditionally or conditioned by given control signals. That is to say, we are interested in the $p(\mathbf{x}|\bm{c})$, where
\begin{equation}
    \bm{c} = \begin{cases}
    \bm{c}_{null}, & \text{unconditional generation}, \\
    \{\bm{c}_n, \bm{c}_p, \bm{c}_c\}, & \text{conditional generation}.
    \end{cases}
\end{equation}
Derived from the reverse-time SDE in Equation~\ref{eq:reverse_sde}, the conditional reverse-time SDE can be described:
\begin{equation}
    \mathrm{d}\mathbf{x}(t) = [\mathbf{f}(\mathbf{x}, t) - g(t)^2 \nabla_{\mathbf{x}}\mathrm{log}p_t(\mathbf{x}|\bm{c})] \mathrm{d} t + g(t) \mathrm{d} \overline{\mathbf{w}},
    \label{eq:cond_reverse_sde}
\end{equation}
where $\overline{\mathbf{w}}$ is a standard Wiener process, $\nabla_{\mathbf{x}}\mathrm{log}p_t(\mathbf{x}|\bm{c})$ is the conditioned noise-perturbed score function, $\mathbf{f}(\cdot,t)$ and $g(\cdot)$ are the drift coefficient and diffusion coefficient of $\mathbf{x}(t)$ respectively.
Thus, by given different $\bm{c}$ for the reverse SDE process, the pianoroll generation stage in \myname{} can achieve unconditional music generation and conditional music generation respectively.

\subsubsection{Fine-grained editing}
\label{sec:controllable_scenario}
Similar to \citet{meng2021sdedit}, the diffusion probabilistic model can serve various fine-grained music editing tasks. We formulate and introduce the following fine-grained music editing tasks, which are illustrated in Figure~\ref{fig:editing_tasks} from top to bottom respectively. All of the following tasks can be implemented with the same algorithm framework (see Algorithm~\ref{alg:vpsde_mask}) with different process operations to obtain the input pianoroll $\mathbf{x}$ and mask $\bm{\Omega}$ and different reverse steps $t_0$. The mask $\bm{\Omega}$ demonstrates the regions that need to be reserved all the time by setting the value to $0$ and the regions that need to be replaced by setting the value to $1$. 
Besides the style transfer which requires control signals, other fine-grained music editing tasks can be conducted both in unconditional and conditional settings.

\begin{wrapfigure}{R}{0.6\textwidth}
\vspace{-0.5cm}
\begin{minipage}{0.6\textwidth}
\begin{algorithm}[H]
\small
\caption{Fine-grained Music Editing (VP-SDE)}
\label{alg:vpsde_mask}
\begin{algorithmic}
    \REQUIRE $\mathbf{x}$ (the input pianoroll), $\bm{\Omega}$ (mask for edited regions), $t_0$ (reverse steps, SDE hyper-parameter), $N$ (diffusion steps), $K$ (total repeats)
    \STATE  $\Delta t \gets \frac{t_0}{N}$
    \STATE $\mathbf{x}_0 \gets \mathbf{x}$ 
    \STATE $\alpha(t_0) \gets \prod_{i=1}^{N}(1-\beta(\frac{it_0}{N})\Delta t)$
    \FOR{$k \gets 1$ \textbf{to} $K$}
    \STATE $\mathbf{z} \sim \mathcal{N}(\mathbf{0}, \bm{\mathit{I}})$
    \STATE $\mathbf{x} \gets \Big [(\mathbf{1} - \bm{\Omega}) \odot \sqrt{\alpha(t_0)}\ \mathbf{x}_0 + \bm{\Omega} \odot \sqrt{\alpha(t_0)}\ \mathbf{x} + \sqrt{1-\alpha(t_0)}\ \mathbf{z} \Big ]$
    \FOR{$n \gets N$ \textbf{to} $1$}
    \STATE $t \gets t_0\frac{n}{N}$
    \STATE $\mathbf{z} \sim \mathcal{N}(\mathbf{0}, \bm{\mathit{I}})$
    \STATE $\alpha(t) \gets \prod_{i=1}^{n}(1-\beta(\frac{it_0}{N})\Delta t)$
    \STATE 
    $\mathbf{x} \gets \Big \{(\mathbf{1} - \bm{\Omega}) \odot \Big ( \sqrt{\alpha(t)}\ \mathbf{x}_0 + \sqrt{1-\alpha(t)}\ \mathbf{z} \Big ) + \bm{\Omega} \odot \Big [ \frac{1}{\sqrt{1-\beta(t)\Delta t}} \Big (\mathbf{x} + \beta(t)\Delta t \bm{s_{\theta}}(\mathbf{x}, t)  + \sqrt{\beta(t)\Delta t}\ \mathbf{z} \Big ) \Big ] \Big \}$ 
    \ENDFOR
    \ENDFOR
    \RETURN $\mathbf{x}$
\end{algorithmic}
\end{algorithm}
\end{minipage}
\vspace{-0.5cm}
\end{wrapfigure}

% \begin{algorithm}[!t]
% \caption{Fine-grained Music Editing (VP-SDE)}
% \label{alg:vpsde_mask}
% \small
% \begin{algorithmic}[1]
%     \REQUIRE $\mathbf{x}$ (the input pianoroll), $\bm{\Omega}$ (mask for edited regions), $t_0$ (reverse steps, SDE hyper-parameter), $N$ (diffusion steps), $K$ (total repeats)
%     \STATE  $\Delta t \gets \frac{t_0}{N}$
%     \STATE $\mathbf{x}_0 \gets \mathbf{x}$ 
%     \STATE $\alpha(t_0) \gets \prod_{i=1}^{N}(1-\beta(\frac{i}{N}t_0)\Delta t)$
%     \FOR{$k \gets 1$ \textbf{to} $K$}
%     \STATE $\mathbf{z} \sim \mathcal{N}(\mathbf{0}, \bm{\mathit{I}})$
%     \STATE $\mathbf{x} \gets \left [(\mathbf{1} - \bm{\Omega}) \odot \sqrt{\alpha(t_0)}\ \mathbf{x}_0 + \bm{\Omega} \odot \sqrt{\alpha(t_0)}\ \mathbf{x}  + \sqrt{1-\alpha(t_0)}\ \mathbf{z} \right]$
%     \FOR{$n \gets N$ \textbf{downto} $1$}
%     \STATE $t \gets t_0\frac{n}{N}$
%     \STATE $\mathbf{z} \sim \mathcal{N}(\mathbf{0}, \bm{\mathit{I}})$
%     \STATE $\alpha(t) \gets \prod_{i=1}^{n}(1-\beta(\frac{i}{N}t_0)\Delta t)$
%     \STATE 
%     \begin{aligned}
%     $\mathbf{x} \gets & \left \{(\mathbf{1} - \bm{\Omega}) \odot \left ( \sqrt{\alpha(t)}\ \mathbf{x}_0 + \sqrt{1-\alpha(t)}\ \mathbf{z} \right ) + \bm{\Omega} \odot \left [ \frac{1}{\sqrt{1-\beta(t)\Delta t}}  \right \right \\ & \left \left \left (\mathbf{x} + \beta(t)\Delta t \bm{s_{\theta}}(\mathbf{x}, t)  + \sqrt{\beta(t)\Delta t}\ \mathbf{z} \right ) \right ] \right \}$ 
%     \end{aligned}
%     \ENDFOR
%     \ENDFOR
%     \RETURN $\mathbf{x}$
    
% \end{algorithmic}
% \end{algorithm}

\noindent\textbf{Stroke-based generation.} Like stroke-based image generation mentioned in~\citet{meng2021sdedit}, stroke-based generation aims to generate the realistic pianoroll with the given stroke pianoroll $\mathbf{x}$. The generated pianorolls are expected to balance between faithfulness and realism, which means that they should not only be similar to the given stroke pianoroll but also be authentic and reasonable. One can draw a stroke pianoroll with a specific structure, thus enabling structure-conditioned music generation. The mask $\bm{\Omega}$ is set to $1$ for all regions and the reverse steps $t_0$ is set to $0.4$. 

\noindent\textbf{Stroke-based editing.} When someone is not satisfied with an existing pianoroll and wants to edit a certain part of it, the stroke-based editing can help. Given a pianoroll with stroke edits $\mathbf{x}$, we can generate a realistic pianoroll that follows the editing information, keeping the other parts from being changed. Stroke-based editing allows users to polish a given pianoroll to their liking, which is useful for eliminating bad cases and personalized music generation. The mask $\bm{\Omega}$ is set to $1$ only for the edited regions and the reverse steps $t_0$ is set to $0.4$. 

\noindent\textbf{Inpainting/outpainting.} Similar to image inpainting~\citep{yeh2017semantic} and outpainting~\citep{wang2014biggerpicture}, we would like to reconstruct missing regions or extend the border of existing pianoroll $\mathbf{x}$. Inpainting can be used for music detail filling and outpainting can be used for music continuation, which is important to generate music pieces with flexible lengths. For inpainting, the mask $\bm{\Omega}$ is set to $1$ for the missing regions and the reverse steps $t_0$ is set to $1$. And for outpainting, we concatenate the pianoroll and a random initialized part as input $\mathbf{x}$ and set the mask $\bm{\Omega}$ to $1$ only for the randomly initialized regions and the reverse steps $t_0$ to $1$.

\noindent\textbf{Combination.} Another important scenario is combining several music segments together harmoniously, which can be applied when someone likes several segments and wants them to appear in the same musical piece. We concatenate the pieces with some parts which are sampled from the prior distribution $p_T$, and use this as input $\mathbf{x}$. Similar to outpainting, the mask $\bm{\Omega}$ is set to $1$ only for the randomly initialized regions and the number of reverse steps $t_0$ is set to $1$.

\noindent\textbf{Style transfer.} As described in~\citet{wu2021musemorphose}, given an existing musical piece $\mathbf{x}$, we can change it to another style by adjusting some control signals such as rhythmic intensity, note density, pitch distribution, etc. For example, if the note density of a certain music piece increases, the music piece will sound more intense or upbeat. We can achieve precise local control cause control signals in \myname{} are fine-grained. The mask $\bm{\Omega}$ is set to $1$ for all regions and $t_0$ is set to $0.4$.  

\subsection{MIDI-event generation stage}
\label{sec:performance_gen}
The MIDI-event generation stage is designed for refining the generated pianorolls from the prior stage with more precise music performance details by generating the final MIDI-event sequence auto-regressively, thus being able to benefit from the advantages of MIDI-event representation for music generation. 
This stage contains a score encoder to encode music scores (pianoroll) into score condition and an auto-regressive decoder to generate MIDI-event tokens step by step.
The score encoder (see Figure~\ref{fig:score_encoder} in Appendix~\ref{app:model_arch_details}) is a 12-layer convolution network with group normalization, which takes the synthesized pianorolls as input to generate the corresponding score condition. The score condition is then concatenated with \textit{barpos embedding}, which is used to indicate the position and introduced in \citet{ren2020popmag}, and tiled to the same length with decoder input (MIDI-event token sequence) according to the bar information from decoder input. We call the set of these operations as ``\textit{expand by bar}" and illustrate it in Figure~\ref{fig:add_score_condition} in Appendix~\ref{app:model_arch_details}. Finally, the score condition can be added to the decoder input and forwarded to the auto-regressive decoder. The auto-regressive decoder is a transformer decoder~\citep{vaswani2017attention,dai2019transformer}, which can also be regarded as a conditional language model, helping with eliminating outliers predicted in pianorolls and adding more precise music performance details.

\section{Experiments}
\label{sec:exp}
In this section, we first introduce the experimental setup including dataset, evaluation metrics, baselines, etc. Then we report the results of unconditioned and conditioned generation with \myname{}. And finally, we show the performance of \myname{} in various fine-grained music editing tasks with evaluation results and cases. The audio samples can be found in our demo page~\footnote{https://SDMuse.github.io/posts/sdmuse/}.

\subsection{Experimental settings}
\label{sec:exp_setting}

\noindent\textbf{Datasets.}
\label{sec:datasets}
In our experiments, we use the \textit{ailabs1k7} dataset introduced by \citet{hsiao2021compound}, which contains 1,748 pieces of pop piano performance. We process all the pieces in training set into 32-bar segments by sliding window, with window size of 32 bar and hopping size of 4 bar, thus obtain around 15,000 segments for the training of conditioned diffusion model in pianoroll generation stage and encoder-decoder in MIDI-event generation stage. For pianoroll, we set the granularity to 1 beat, which means the length of pianoroll $n$ is the beat number of the corresponding musical piece. And for MIDI-event sequence, we follow the representation in \citet{ren2020popmag}, ignoring the track and instrument information because the music of \textit{ailabs1k7} dataset is single-track polyphonic music. 

\noindent\textbf{Evaluation metrics.}
\label{sec:eval_metrics}
We evaluate our results based on \textit{quality} and \textit{controllability}, for which we performed both objective and subjective assessments. As listed in Table~\ref{tab:eval_metrics}, to evaluate \textit{quality}, we use PD and DD scores introduced in \citet{sheng2021songmass} and conduct subjective evaluation to obtain the overall perceptive scores. On the other hand, for quantifying \textit{controllability} of conditioned music generation and editing, we calculate the $L_2$ distance of control signals (CSD) between the given one and those of generated output. Also, when conducting fine-grained music editing like stroke-based generation, we compute the overlap ratio (OR) between the generated pianorolls and the input stroke pianorolls. And we evaluate the consistency of the edited samples in fine-grained music editing subjectively.
Similar to~\citet{zhang2022relyme,guo2022automatic}, we invite 10 participants with music knowledge to give their scores (five-point scales, 1 for bad and 5 for excellent) of randomly selected samples. The detailed instruction given to annotators are posted in Appendix~\ref{app:human_eval}.

\begin{table}[t]
    \small
	\centering
	\vspace{-0.5cm}
	\caption{The evaluation metrics of \myname{} in terms of both \textit{quality} and \textit{controllability}.}
	\begin{tabular}{m{1.5cm} | m{5cm} | m{5cm}}
		\toprule
		 & \multicolumn{1}{c}{\textit{Quality}} & \multicolumn{1}{c}{\textit{Controllability}} \\
		\midrule
		\multirow{2}{*}{Objective} & pitch distribution similarity (PD) & control signal distance (CSD) \\
		 & duration distribution similarity (DD) & overlap ratio (OR) \\ 
        \midrule
		Subjective & overall perceptive score & overall consistency score \\
        \bottomrule
	\end{tabular}
	\label{tab:eval_metrics} 
	\vspace{-0.5cm}
\end{table}

\noindent\textbf{Baselines.}
\label{sec:baselines}
For comparison, we choose different types of symbolic music generation and style transfer models as our baselines: 1) REMI~\citep{huang2020pop}; 2) CPW~\citep{hsiao2021compound} and 3) MuseMorphose~\citep{wu2021musemorphose}. We use the official implementation of each model and train these three models with the same training set. Considering that we are the first one to conduct fine-grained music editing tasks, we only compare the quality of generated outputs with these baselines.

\noindent\textbf{Model configuration.}
\label{sec:model_config}
For the pianoroll generation stage, we use Gaussian diffusion model~\footnote{https://github.com/openai/guided-diffusion} and adjust the UNet architecture~\citep{cciccek20163d} to make sure it can take control signals as condition.
And for the MIDI-event generation stage, we use a 4-layer transformer decoder and 12-layer convolution 1D encoder. 
Other details about the model hyper-parameters are listed in the Appendix~\ref{app:model_config}.

\noindent\textbf{Training setup.}
\label{sec:training_setup}
The training data of both modules is cut into 32-bar segments. We train the diffusion model with diffusion step of 100 and use the linear noise schedule with max beta of 0.02. The training process of the pianoroll generation stage takes about 2 days on 1 A100 GPU with batch size of 32 pianorolls. And the MIDI-event generation stage are trained around 12 hours on 1 A100 GPU with batch size of 2000 MIDI-event tokens.

\subsection{Generation from scratch}
\label{sec:exp_main_res}

\begin{table}[t]
	\centering
    \small
    \vspace{-0.5cm}
	\caption{Objective and subjective results of baseline systems in music generation, and \myname{} in generation from scratch tasks (both unconditioned and conditioned on given control signals) and fine-grained music editing tasks in terms of \textbf{\textit{quality}}. Settings with * notation have high PD, DD and subjective perceptive scores because they are based on existing musical pieces with only minor edits. The overall perceptive scores are calculated with 95\% confidence intervals.}
	% \resizebox{\linewidth}{!}{
	\begin{tabular}{c l c c c}
		\toprule
		\multicolumn{1}{c}{\multirow{2}{*}{Task}} & \multicolumn{1}{c}{\multirow{2}{*}{Model / Setting}} & \multicolumn{2}{c}{Objective} & \multicolumn{1}{c}{Subjective} \\
		\cmidrule(lr){3-4}
		\cmidrule(lr){5-5}
		 & & PD~$\uparrow$ & DD~$\uparrow$ & overall perceptive score~$\uparrow$ \\
		\midrule
		\cellcolor{gray!25} & \cellcolor{gray!25}GT & \cellcolor{gray!25}-- & \cellcolor{gray!25}-- & \cellcolor{gray!25}4.07 ($\pm 0.09$)  \\
		\midrule
		\multirow{4}{*}{Generation} & REMI~\citep{huang2020pop} & 0.82 & 0.76 & 3.52 ($\pm 0.07$)  \\
		 & CPW~\citep{hsiao2021compound} & 0.74 & 0.80 & 3.71 ($\pm 0.06$) \\
        \cmidrule(lr){2-5}
		 & \myname{} (unconditioned) & 0.84 & \textbf{0.81} & 3.69 ($\pm 0.06$) \\
		 & \myname{} (conditioned) & \textbf{0.88} & 0.79 & \textbf{3.72} ($\pm 0.06$) \\
		\midrule
		\multirow{7}{*}{Editing} & MuseMorphose~\citep{wu2021musemorphose}* & 0.68 & 0.81 & 3.80 ($\pm 0.06$) \\
        \cmidrule(lr){2-5}
		 & \myname{} (stroke-based generation) & 0.84 & 0.80 & 3.47 ($\pm 0.07$) \\
		 & \myname{} (stroke-based editing)* & 0.96 & 0.88 & 3.81 ($\pm 0.07$) \\
		 & \myname{} (inpainting)* & 0.96 & 0.87 & 3.70 ($\pm 0.06$) \\
		 & \myname{} (outpainting) & 0.79 & 0.75 & 3.63 ($\pm 0.07$) \\
		 & \myname{} (combination) & 0.86 & 0.83 & 3.59 ($\pm 0.09$) \\
		 & \myname{} (style transfer)* & 0.92 & 0.80 & 3.77 ($\pm 0.07$) \\
        \bottomrule
	\end{tabular}
	% }
	\label{tab:quality_res}
\end{table}

\begin{figure}[!t]
	\centering
	\begin{subfigure}[t]{0.24\textwidth}
	    \captionsetup{justification=centering}
		\centering
	    \includegraphics[width=\textwidth,trim={0cm 0.5cm 0cm 0.5cm}, clip=true]{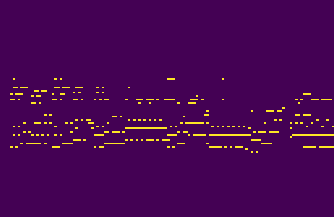}
	    \includegraphics[width=\textwidth,trim={0cm 0.5cm 0cm 0.5cm}, clip=true]{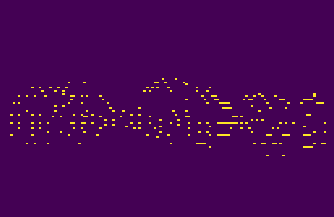}
	    \includegraphics[width=\textwidth,trim={0cm 0.5cm 0cm 0.5cm}, clip=true]{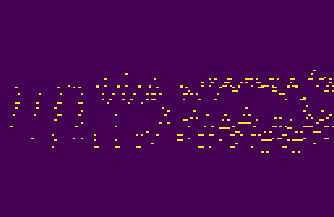}
	    \subcaption{REMI}
	\end{subfigure}
	\begin{subfigure}[t]{0.24\textwidth}
	    \captionsetup{justification=centering}
		\centering
	    \includegraphics[width=\textwidth,trim={0cm 0.5cm 0cm 0.5cm}, clip=true]{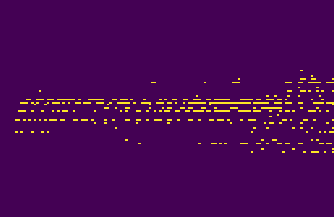}
	    \includegraphics[width=\textwidth,trim={0cm 0.5cm 0cm 0.5cm}, clip=true]{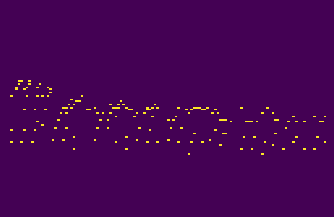}
	    \includegraphics[width=\textwidth,trim={0cm 0.5cm 0cm 0.5cm}, clip=true]{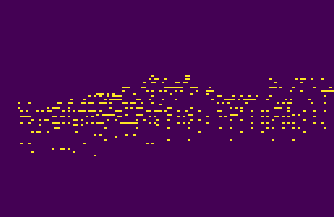}
	    \subcaption{CPW}
	\end{subfigure}
	\begin{subfigure}[t]{0.24\textwidth}
	    \captionsetup{justification=centering}
		\centering
	    \includegraphics[width=\textwidth,trim={0cm 0.5cm 0cm 0.5cm}, clip=true]{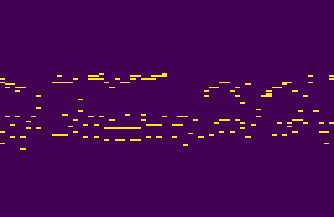}
	    \includegraphics[width=\textwidth,trim={0cm 0.5cm 0cm 0.5cm}, clip=true]{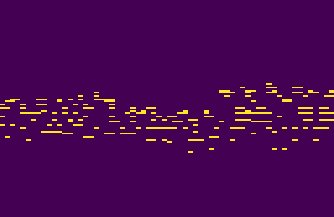}
	    \includegraphics[width=\textwidth,trim={0cm 0.5cm 0cm 0.5cm}, clip=true]{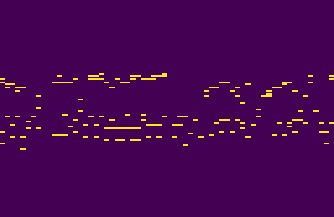}
	    \subcaption{MuseMorphose}
	\end{subfigure}
	\begin{subfigure}[t]{0.24\textwidth}
	    \captionsetup{justification=centering}
		\centering
	    \includegraphics[width=\textwidth,trim={0cm 0.5cm 0cm 0.5cm}, clip=true]{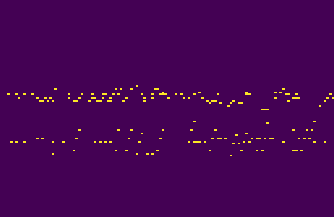}
	    \includegraphics[width=\textwidth,trim={0cm 0.5cm 0cm 0.5cm}, clip=true]{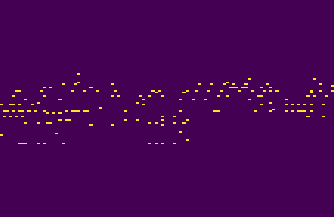}
	    \includegraphics[width=\textwidth,trim={0cm 0.5cm 0cm 0.5cm}, clip=true]{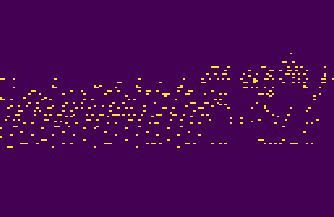}
	    \caption{\myname{} (Ours)}
	\end{subfigure}
	\caption{The pianorolls extracted from the music generated by \myname{} and baseline systems.}
	\label{fig:uncond_figs}
\end{figure}
\subsubsection{Unconditioned generation}
We first evaluate the performance of \myname{} on the unconditioned music generation task by just setting the control signals $\bm{c}$ as $\bm{c}_{null}$ in the pianoroll generation stage. As shown in Table~\ref{tab:quality_res}, denoted as \myname{} (unconditioned), while unconditioned generation is not our primary goal, we find that \myname{} achieves comparable results to the state-of-the-art music generation models, which indicates the effectiveness of our training strategy of diffusion probabilistic model that switches between unconditioned and conditioned generation settings.
We present qualitative comparison results in Figure~\ref{fig:uncond_figs} by just showing the pianorolls extracted from the final outputs of \myname{} and baselines.

\subsubsection{Conditioned generation}
In order to assess the performance of \myname{} when generating musical pieces from scratch conditioned on given control signals, we use the control signals extracted from the test set during the pianoroll generation stage. 
The \textit{quality} results are also listed in Table~\ref{tab:quality_res}, denoted as \myname{} (conditioned), demonstrating that with the guidance of explicit music information from control signals, \myname{} can obtain better generation quality with reasonable listening experience compared to unconditioned generation.
And the \textit{controllability} results are presented in Table~\ref{tab:controllability_res}, indicating that \myname{} has the ability to generate musical pieces based on the control signals faithfully. 
Also, as a complement, we compare the note density $\bm{c}_n$ and the pitch distribution $\bm{c}_p$ extracted from output music with the given ones, illustrated in Figure~\ref{fig:control_signals_word_emb}, for a visual demonstration of the faithfulness in terms of control signals during conditioned generation.

\subsection{Fine-grained editing}
\begin{table}[t]
    \small
	\centering
	\vspace{-0.5cm}
	\caption{Objective and subjective results of style transfer baseline and \myname{} in conditioned music generation task and various fine-grained music editing tasks in terms of \textbf{\textit{controllability}}. The subjective scores are calculated with 95\% confidence intervals.}
	\resizebox{\linewidth}{!}{
	\begin{tabular}{l c c c c}
		\toprule
		\multicolumn{1}{c}{\multirow{2}{*}{Model / Setting}} & \multicolumn{3}{c}{Objective} & \multicolumn{1}{c}{Subjective} \\
		\cmidrule(lr){2-4}
		\cmidrule(lr){5-5}
		 & CSD ($\bm{n}_n$)~$\downarrow$ & CSD ($\bm{n}_p$)~$\downarrow$ & OR $\uparrow$ & overall consistency score~$\uparrow$ \\
		\midrule
		MuseMorphose~\citep{wu2021musemorphose} & -- & -- & -- & 3.87 ($\pm 0.07$) \\
		\midrule
		\myname{} (conditioned) & 0.06 & 0.15 & -- & 3.77 ($\pm 0.06$) \\
		\midrule
		\myname{} (stroke-based generation) & -- & -- & 0.85 & 3.43 ($\pm 0.10$) \\
		\myname{} (stroke-based editing) & -- & -- & 0.81 & 3.89 ($\pm 0.08$) \\
% 		\myname{} (inpainting) & 0.0 & 0.0 & 0.0 ($\pm 0.00$) \\
% 		\myname{} (outpainting) & 0.0 & 0.0 & 0.0 ($\pm 0.00$) \\
% 		\myname{} (combination) & 0.0 & 0.0 & 0.0 ($\pm 0.00$) \\
		\myname{} (style transfer) & 0.12 & 0.38 & -- & \textbf{4.02} ($\pm 0.06$) \\
        \bottomrule
	\end{tabular}
	}
	\label{tab:controllability_res}
\end{table}
\begin{figure}[!t]
	\centering
	\begin{subfigure}[t]{0.19\textwidth}
	    \captionsetup{justification=centering}
		\centering
	    \begin{minipage}{0.96\textwidth}
		    \includegraphics[width=\textwidth,trim={0cm 0cm 0cm 0cm}, clip=true]{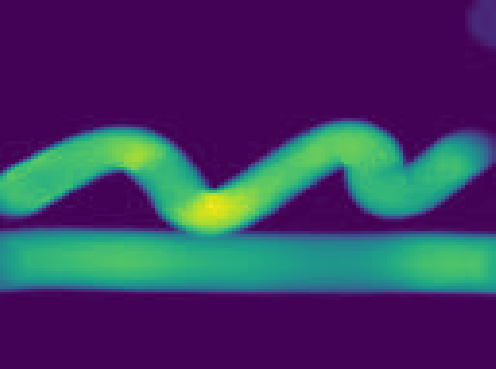}
		\end{minipage}
	    \subcaption{Input.}
	\end{subfigure}
	\begin{subfigure}[t]{0.38\textwidth}
	    \captionsetup{justification=centering}
		\centering
		\begin{minipage}{0.48\textwidth}
		    \includegraphics[width=\textwidth,trim={0cm 0cm 0cm 0cm}, clip=true]{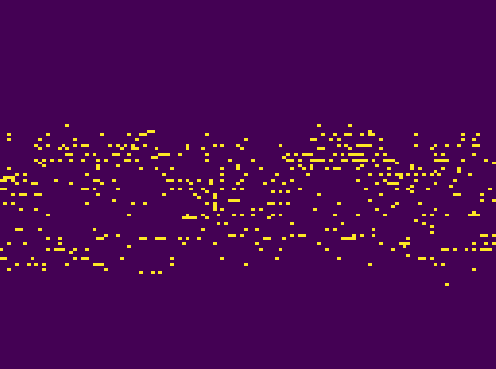}
		\end{minipage}
		\begin{minipage}{0.48\textwidth}
		    \includegraphics[width=\textwidth,trim={0cm 0cm 0cm 0cm}, clip=true]{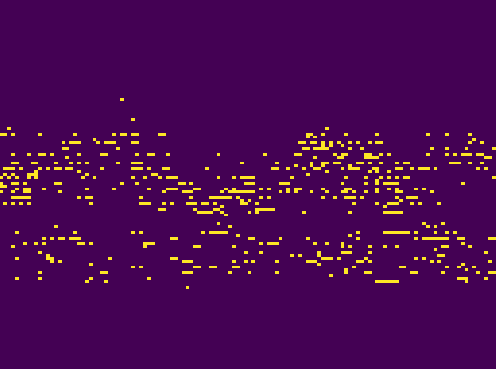}
		\end{minipage}
	    \subcaption{Generated pianorolls.}
	\end{subfigure}
	\begin{subfigure}[t]{0.38\textwidth}
	    \captionsetup{justification=centering}
		\centering
	    \begin{minipage}{0.48\textwidth}
		    \includegraphics[width=\textwidth,trim={0cm 0cm 0cm 0cm}, clip=true]{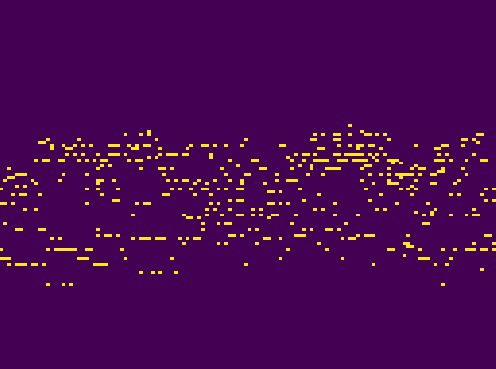}
		\end{minipage}
		\begin{minipage}{0.48\textwidth}
		    \includegraphics[width=\textwidth,trim={0cm 0cm 0cm 0cm}, clip=true]{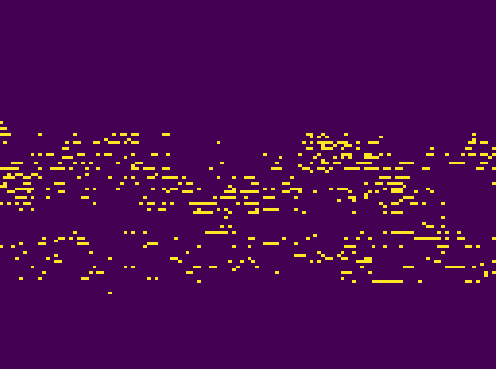}
		\end{minipage}
	    \caption{Extracted pianorolls of final outputs.}
	\end{subfigure}
	\caption{The pianorolls generated by the pianoroll generation stage and extracted from the final output music generated by MIDI-event generation stage in stroke-based generation task.}
	\label{fig:stroke_generation}
\end{figure}

% \begin{figure}[!t]
% 	\centering
% 	\begin{subfigure}[t]{0.19\textwidth}
% 	    \captionsetup{justification=centering}
% 		\centering
% 	    \begin{minipage}{0.96\textwidth}
% 		    \includegraphics[width=\textwidth,trim={0cm 0cm 0cm 0cm}, clip=true]{figs/stroke/stroke_input.pdf}
% 		\end{minipage}
% 	    \subcaption{Input.}
% 	\end{subfigure}
% 	\begin{subfigure}[t]{0.38\textwidth}
% 	    \captionsetup{justification=centering}
% 		\centering
% 		\begin{minipage}{0.48\textwidth}
% 		    \includegraphics[width=\textwidth,trim={0cm 0cm 0cm 0cm}, clip=true]{figs/stroke/stroke_cond_1.pdf}
% 		\end{minipage}
% 		\begin{minipage}{0.48\textwidth}
% 		    \includegraphics[width=\textwidth,trim={0cm 0cm 0cm 0cm}, clip=true]{figs/stroke/stroke_cond_2.pdf}
% 		\end{minipage}
% 	    \subcaption{Generated pianorolls.}
% 	\end{subfigure}
% 	\begin{subfigure}[t]{0.38\textwidth}
% 	    \captionsetup{justification=centering}
% 		\centering
% 	    \begin{minipage}{0.48\textwidth}
% 		    \includegraphics[width=\textwidth,trim={0cm 0cm 0cm 0cm}, clip=true]{figs/stroke/stroke_pred_1.pdf}
% 		\end{minipage}
% 		\begin{minipage}{0.48\textwidth}
% 		    \includegraphics[width=\textwidth,trim={0cm 0cm 0cm 0cm}, clip=true]{figs/stroke/stroke_pred_2.pdf}
% 		\end{minipage}
% 	    \caption{Extracted pianorolls of final outputs.}
% 	\end{subfigure}
% 	\caption{The pianorolls generated by the pianoroll generation stage and extracted from the final output music generated by MIDI-event generation stage in stroke-based generation task.}
% 	\label{fig:stroke_generation}
% \end{figure}
For aforementioned fine-grained editing tasks (see Section~\ref{sec:controllable_scenario} for details), we edit existing musical pieces in corresponding ways and evaluate these tasks respectively. The \textit{quality} results are shown in Table~\ref{tab:quality_res} and the \textit{controllability} results are shown in Table~\ref{tab:controllability_res}. 
It is obvious that the tasks with only minor edits, such as stroke-based editing, inpainting and style transfer, perform well in both \textit{quality} and \textit{controllability}. For the stroke-based generation which highly depends on the given stroke pianorolls, there is a trade-off between the PD/DD scores and OR score.
Figure~\ref{fig:stroke_generation} illustrates the process of stroke-based generation task, including the input stroke pianoroll $\mathbf{x}$, the output of pianoroll generation stage, and the pianorolls extracted from the final MIDI-event sequence. Illustrations of other fine-grained editing tasks and the final musical pieces can be found in our demo page~\footnote{https://\myname.github.io/posts/sdmuse/}.
% The implementation details of each scenario and the qualitative comparison results can be found in the supplementary materials. Both objective and subjective evaluation results verify the effectiveness of \myname{} in terms of low-level controllability on diverse tasks.

\subsection{Method analyses}
\label{sec:analyses}
We conduct more explorations on \myname{} and put the results in Appendix~\ref{app:method_analyses} due to the limited space, including: 1) the refinement performance of MIDI-event generation stage to verify the ability of eliminating outliers and adding music performance details; 2) the comparison among different embedding ways of control signals. In summary, it is observed that MIDI-event generation stage is good at refining pianorolls with outliers. And when involving control signals in conditioned diffusion model, word embedding and direct embedding outperform positional embedding.
% ; 3) the comparison among different input types of MIDI-event generation stage; and 4) the reverse step $t_0$ in some of fine-grained editing tasks.

% \noindent\textbf{Reconstruction Performance} Here we illustrate the reconstruction performance of auto-regressive performance generation module. We compare the onset-roll input of this module and the onset-roll extracted from the output performance as in Figure~\ref{fig:reconstruction}. It is obvious that there are some outliers in the input onset-roll, but after this module, these outliers are eliminated.
% % TODO: 加上对于pianoroll这种数据格式的分析 -> 造成outliers

% \noindent\textbf{Reverse Step}

% \noindent\textbf{Tile-concatenate v.s. Cross-attn}

% \noindent\textbf{Onset-roll v.s. Onset Tokens}

% \noindent\textbf{Embedding Ways of Control Signals} As shown in Figure~\ref{fig:embedding_ways}

\section{Conclusion}
\label{sec:conclusion}
In this paper, we propose \myname{}, a unified \underline{S}tochastic \underline{D}ifferential \underline{Mus}ic \underline{e}diting and generation framework via hybrid representations. \myname{} can not only compose whole musical pieces from scratch (both unconditionally and conditioned on given control signals), but also edit existing musical pieces in different ways according to various demands. As two different symbolic music representations, pianoroll is more appropriate for extracting and editing perceptive music information, such as structure, while MIDI-event is more ideal for generating and modeling music performance details. Thus, \myname{} contains pianoroll and MIDI-event generation stages to take advantage of hybrid representations. The first stage is based on a diffusion model generative prior and synthesizes or edits pianorolls by iteratively denoising through SDE. And the second stage refines pianorolls with music performance details by generating MIDI-event sequences auto-regressively. Objective and subjective results on \textit{ailabs1k7} dataset demonstrate the effectiveness of our proposed stochastic differential music editing/generation process and hybrid representations. 
% a song-level symbolic music generation model with high quality and different-level controllability. \myname{} consists of two modules: 1) a diffusion-based controllable score generation module that generates music score in onset-roll format according to control signals; 2) an auto-regressive performance generation module that generates music performance in MIDI event token format from the output music score of the prior module. We explore the characteristics of two symbolic music representations and take their advantages respectively in \myname{}. \myname{} is designed for enabling a series of controllable generation scenarios, which helps to achieve flexible collaboration between humans and models.
% Experiments on \textit{ailabs1k7} dataset are evaluated in terms of \textit{quality} and \textit{controllability}. Both objective and subjective results demonstrates the effectiveness of \myname{} in ensuring high quality and different-level controllability at the same time.
In the future, we plan to deploy \myname{} as an interactive website to make it accessible to more people who are interested in it, as well as extend it to other music genres. 

\bibliography{iclr2023_conference}
\bibliographystyle{iclr2023_conference}

\appendix
\newpage
\section{Symbolic music representation}
\label{app:music_repr}

MIDI-event and pianoroll are two of the most common music representations in symbolic music generation works. As listed in Table~\ref{tab:music_repr}, MIDI-event and pianoroll have their own advantages and disadvantages when representing a piece of symbolic music.
An pianoroll (see Figure~\ref{fig:pianoroll}) is like an image, with the horizontal axis representing time and the vertical axis representing pitch. It is closer to the way how humans perceive musical pieces and contains more prior music information. Pitch, duration, and relative position of notes are directly shown on it, so one can easily perceive the music structure, note density, etc. of an entire song when given a pianoroll, which makes it more suitable to be edited by humans. 

Though the MIDI-event (see Figure~\ref{fig:midi_event}) contains all information in pianoroll, when converting the MIDI-event tokens into embeddings whose weights are randomly initialized, the correlations between different MIDI-events are lost and need to be relearned from the training data. Given the circumstance that the amount of high-quality symbolic music training data is limited, it is relatively easier for a deep learning model to extract musical information from pianoroll than MIDI-event.
However, besides MuseGAN \citep{dong2018musegan} that utilizes pianorolls, most symbolic music generation works~\citep{huang2018music,huang2020pop,zhang2022relyme} are trained with MIDI-event sequences. This is due to MIDI-event sequence can carry more precise details than pianoroll, such as velocity, which can enhance the richness and expressiveness of the generated music performance. Furthermore, there is no difference in the way that a pianoroll treats onset and other positions of a note, which is not consistent with real music. MIDI-event, however, regards a note as the unit and is more robust to generate musical pieces.

\section{Model details}

\subsection{Details of model architecture}
\label{app:model_arch_details}
\begin{figure}[!t]
	\centering
	\begin{subfigure}[b]{0.36\textwidth}
	    \captionsetup{justification=centering}
		\centering
	    \includegraphics[width=\textwidth,trim={0cm 0cm 0cm 0cm}, clip=true]{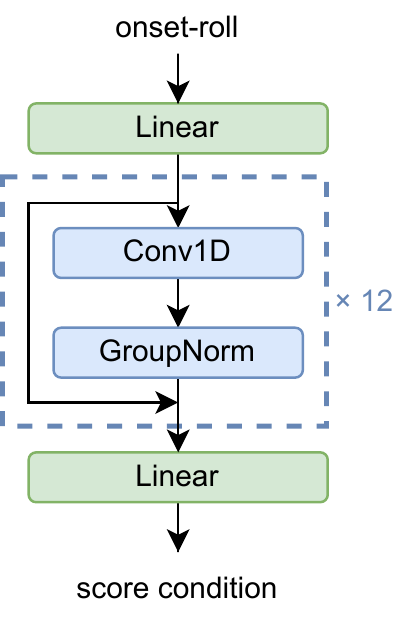}
	    \caption{The structure of score encoder.}
	    \label{fig:score_encoder}
	\end{subfigure}
	\begin{subfigure}[b]{0.54\textwidth}
	    \captionsetup{justification=centering}
		\centering
	    \includegraphics[width=\textwidth,trim={0cm 0cm 0cm 0cm}, clip=true]{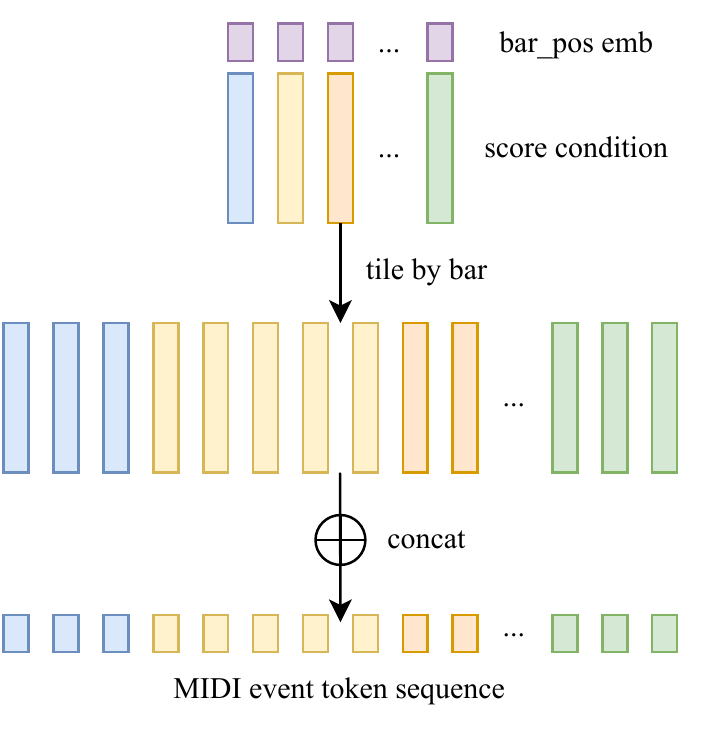}
	    \caption{An illustration of \textit{expand\_by\_bar} operator.}
	    \label{fig:add_score_condition}
	\end{subfigure}
	\caption{The details of score encoder and \textit{expand\_by\_bar} operator in MIDI-event generation stage.}
\end{figure}
As shown in Figure~\ref{fig:score_encoder}, the score encoder of the MIDI-event generation stage in \myname{} has a simple architecture: a 12-layer convolution network with group normalization, as well as two linear layers to process input pianoroll and output score condition. 
And the detailed illustration of \textit{expand\_by\_bar} operator is in Figure~\ref{fig:add_score_condition}. The score condition is then concatenated with \textit{barpos embedding}, which is used to indicate the position and introduced in \citet{ren2020popmag}, and tiled to the same length with decoder input (MIDI-event token sequence) according to the bar information from decoder input. At last, the expanded score condition is add to decoder input.

\subsection{Fine-grained control signals}
\label{app:control_signals}
Denote the processed pianoroll (onsetroll) as  $X_{o} \in \{0, 1\}^{m \times n}$, where $m$ means the number of pitch and $n$ represents the length of pianoroll. Here we design three fine-grained control signals to provide precise controllability over the entire piece. 
\begin{itemize}
    \item Note density: for each piece of music, we can extract the note density vector $\bm{c}_n \in [0, 127]^{1 \times n}$, which indicates how many onsets occur concurrently at each timestep. 
    \item Pitch distribution: the pitch distribution $\bm{c}_p \in [0, 127]^{m \times 1}$ is a vector that represents the distribution of note pitch over the whole musical piece. Specifically, the value in $\bm{c}_p$ of position $p$ is the number of notes whose pitch is $p$.
    \item Chord progression: we extract the chord progression sequence $\bm{c}_c \in [0, 96]^{1 \times n}$ from the original musical piece, where $96$ is the number of chord type.
\end{itemize}
In order to involve these fine-grained control signals into diffusion, we first convert $\bm{c}_n, \bm{c}_p, \bm{c}_c$ to embeddings $\bm{e}_n, \bm{e}_p, \bm{e}_c$ and tile them to the same shape of $m \times n$, which are then concatenated to $x_t$ (the data at t step).

\subsection{Onsetroll}
\label{app:onsetroll}
Denote the original pianoroll in the dataset as $X_{p} \in \{0, 1\}^{m \times n}$, where $m$ means the number of pitch, $n$ means the length of pianoroll and the value is set to 1 when the corresponding position is belong to a note otherwise set to 0. As mentioned in Section~\ref{sec:bg_music_repr}, onset and other positions of the note are treated non-differently in pianoroll, however, in real music, onset is more important than other positions and it directly affects the listening experience. If an extra onset is predicted, it indicates an extra note, while if an extra other position is predicted, it just represents the corresponding note becomes slightly longer. Thus, to grasp the factor that takes care of most, we process the original pianoroll $X_p$ to onsetroll $X_o$ by only keeping the onset information and discarding duration information of each note. Specifically, only the value of the onset position in the pianoroll will remain 1 and other positions of notes will be set to 0. 

\subsection{Model configuration}
\label{app:model_config}
\subsubsection{Diffusion probabilistic model}
Our diffusion probabilistic model is implemented based on the Gaussian diffusion model~\footnote{https://github.com/openai/guided-diffusion}. The size of the input pianoroll is $128 \times 128$ (contains 32 bars and 128 beats). We set the learning rate as $1e-4$, the diffusion step as $100$. For the positive noise function $\beta(t)$ in VP-SDE, we follow~\citet{song2020score,ho2020denoising,dhariwal2021diffusion} and set:
\begin{equation}
    \beta(t) = \beta_{start} + t(\beta_{end} - \beta_{start}),
\end{equation}
where $\beta_{start} = 0.1$ and $\beta_{end} = 20$ in our implementation. And we found that if only using the conditioned diffusion model, $\beta_{end}$ should be smaller, such as $5$ to achieve better performance.
The embedding dim of control signal embeddings $\bm{e}_n$, $\bm{e}_p$, $\bm{e}_c$ is set to $32$ and $p_{uncond}$ is set to $0.5$. 

\subsubsection{Auto-regressive decoder}
In our implementation, the auto-regressive decoder is based on a Transformer decoder with 4 layers. The output of the auto-regressive decoder is split into three parts: pitch, velocity, and duration, and the total loss is calculated by adding the cross entropy losses of these three parts together. The dropout of the auto-regressive decoder is $0.1$, the hidden size is $256$, and the number of heads is $4$.

\section{Subjective evaluation}
\label{app:human_eval}
\begin{figure}
    \centering
    \includegraphics[width=0.8\textwidth,trim={0cm 0cm 0cm 0cm}, clip=true]{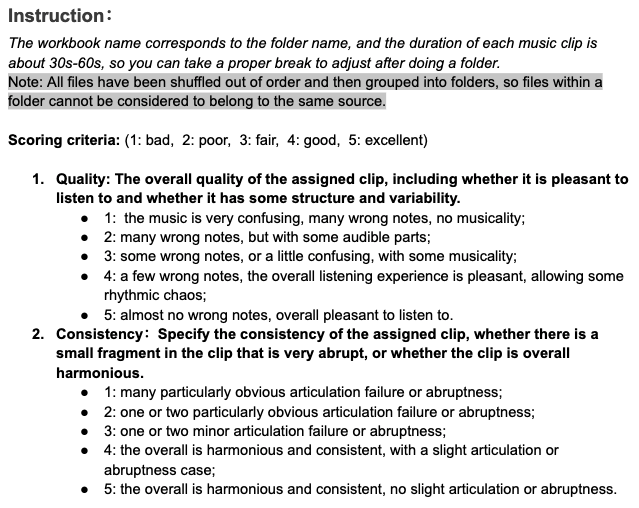}
    \caption{The instructions we give to participants of subjective evaluation part.}
    \label{fig:human_eval}
\end{figure}
We invite 10 people with musical knowledge to give their scores as our subjective evaluation. Here are the instructions we provide to them (Figure~\ref{fig:human_eval}). The five-point scale of subjective evaluation is similar to MOS, which has been widely used in different speech synthesis work~\citep{ren2019fastspeech,zhang2021denoispeech}.

\section{Method analyses}
\label{app:method_analyses}

\subsection{Refinement performance}
\begin{figure}[!t]
	\centering
	\begin{subfigure}[b]{0.4\textwidth}
	    \captionsetup{justification=centering}
		\centering
	    \includegraphics[width=\textwidth,trim={0cm 0cm 0cm 0cm}, clip=true]{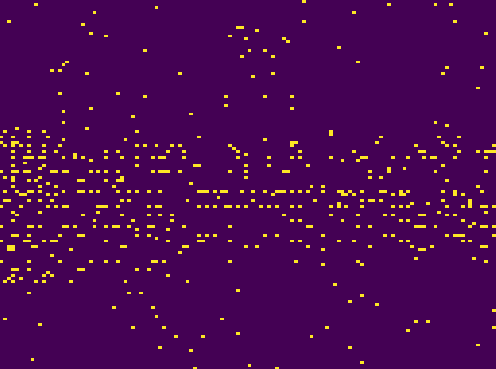}
	    \includegraphics[width=\textwidth,trim={0cm 0cm 0cm 0cm}, clip=true]{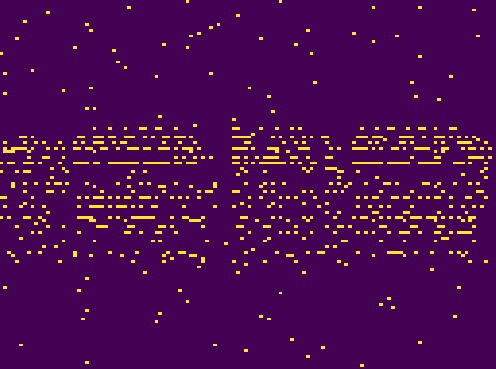}
	    \caption{Input of MIDI-event generation stage.}
	\end{subfigure}
	\begin{subfigure}[b]{0.4\textwidth}
	    \captionsetup{justification=centering}
		\centering
	    \includegraphics[width=\textwidth,trim={0cm 0cm 0cm 0cm}, clip=true]{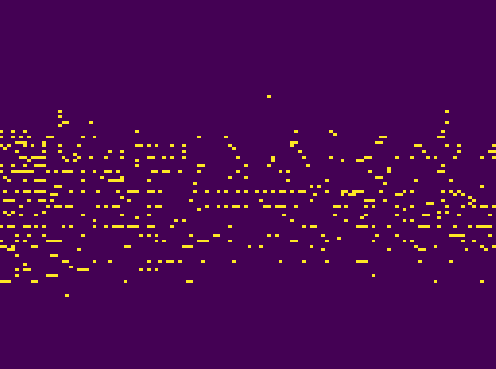}
	    \includegraphics[width=\textwidth,trim={0cm 0cm 0cm 0cm}, clip=true]{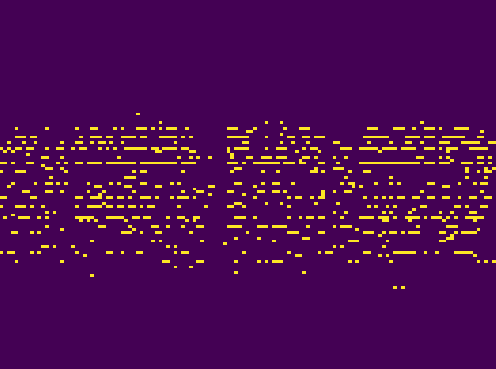}
	    \caption{Output of MIDI-event generation stage.}
	\end{subfigure}
	\caption{The input pianorolls of MIDI-event generation stage and the pianorolls extracted from the output MIDI-event sequences. We highlight some of the eliminated outliers with red boxes for better identify.}
	\label{fig:refinement}
\end{figure}
Here we illustrate the refinement performance of the MIDI-event generation stage. We compare the pianorolls input of this stage and the pianorolls extracted from the output MIDI-event sequences as in Figure~\ref{fig:refinement}. We manually added outliers to the input pianoroll to explore the ability of \myname{} in eliminating outliers. As shown in Figure~\ref{fig:refinement}, these added outliers are removed after the MIDI-event generation stage. Also, the positions of some notes are refined in the MIDI-event sequence generation stage auto-regressively. For a more intuitive listening experience, please refer to our demo page, where we post the audio synthesized from the output of two stages respectively for comparison.

\subsection{Embedding ways of control signals}
\begin{figure}
    \centering
    \begin{subfigure}[t]{0.32\textwidth}
	    \captionsetup{justification=centering}
		\centering
	    \includegraphics[width=\textwidth,trim={0cm 0cm 0cm 0cm}, clip=true]{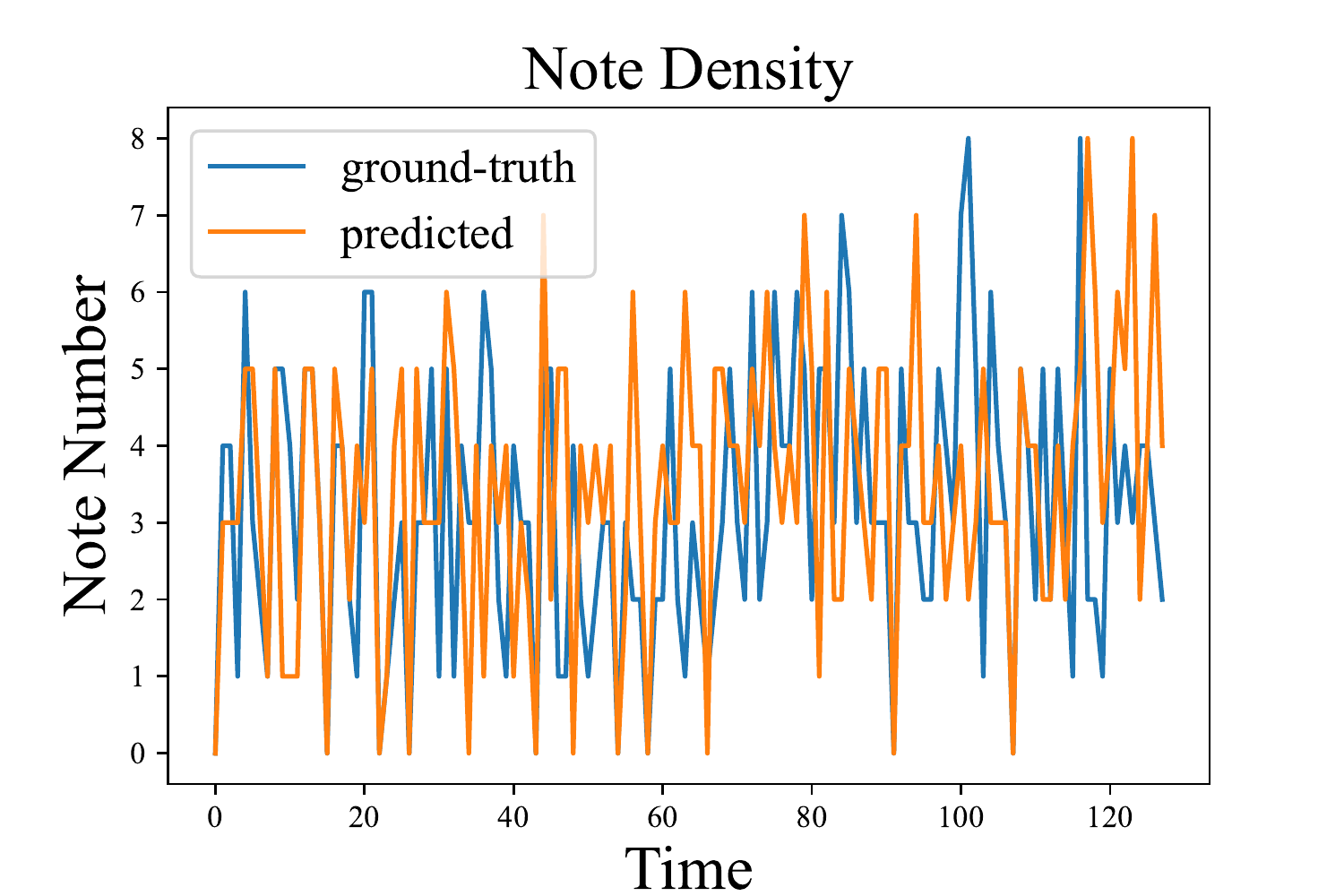}
	    \includegraphics[width=\textwidth,trim={0cm 0cm 0cm 0cm}, clip=true]{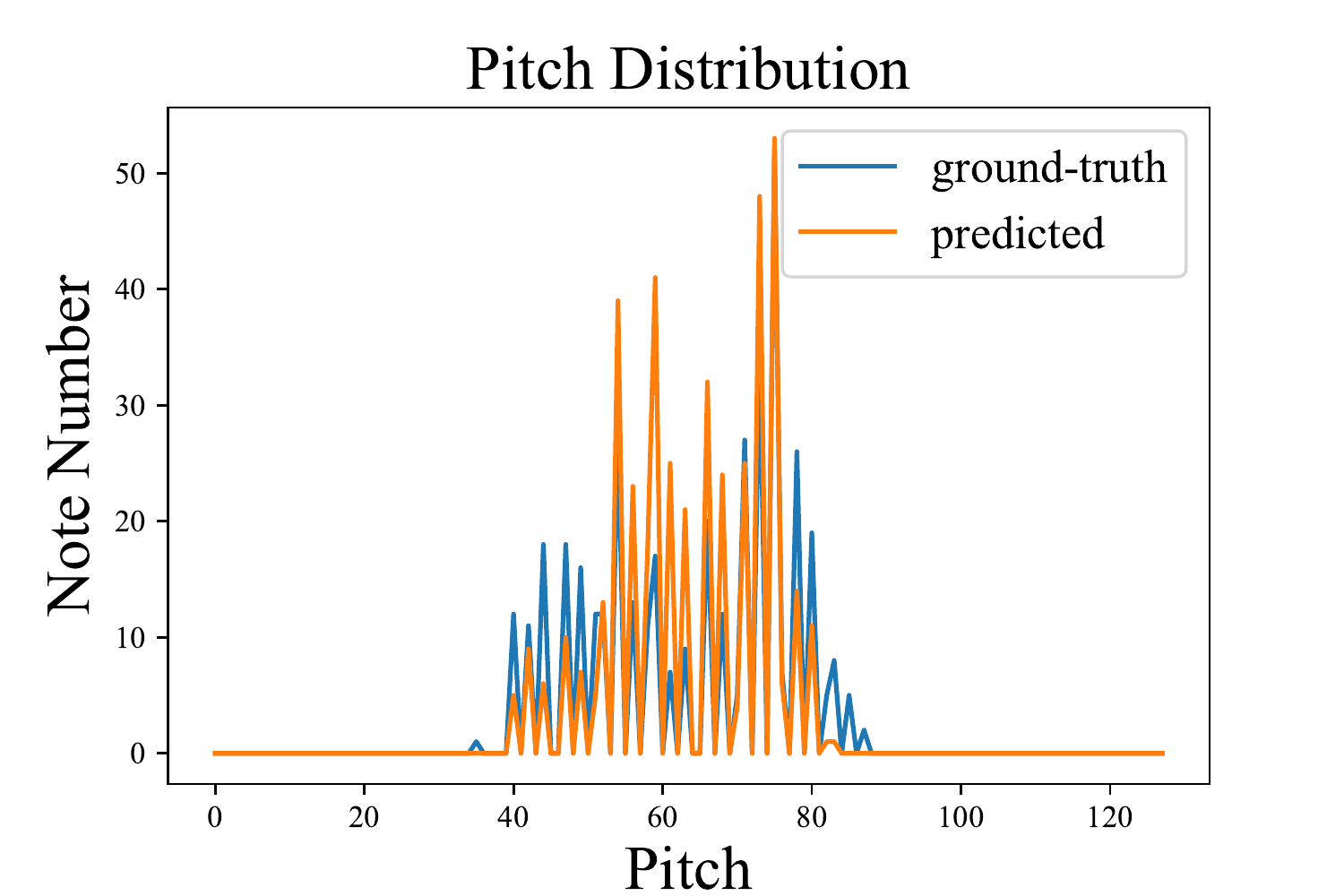}
	    \subcaption{Positional embedding.}
	\end{subfigure}
	\begin{subfigure}[t]{0.32\textwidth}
	    \captionsetup{justification=centering}
		\centering
	    \includegraphics[width=\textwidth,trim={0cm 0cm 0cm 0cm}, clip=true]{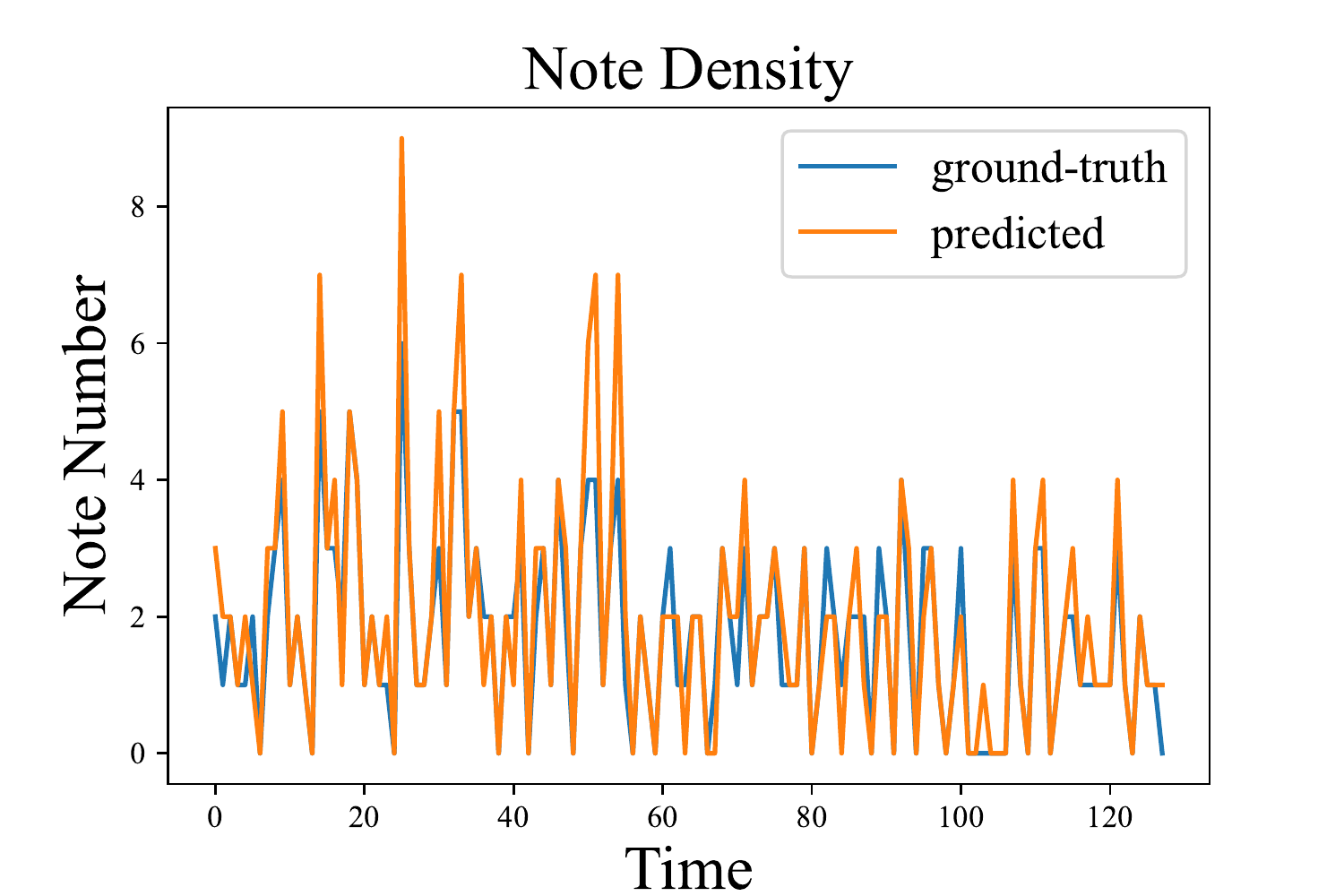}
	    \includegraphics[width=\textwidth,trim={0cm 0cm 0cm 0cm}, clip=true]{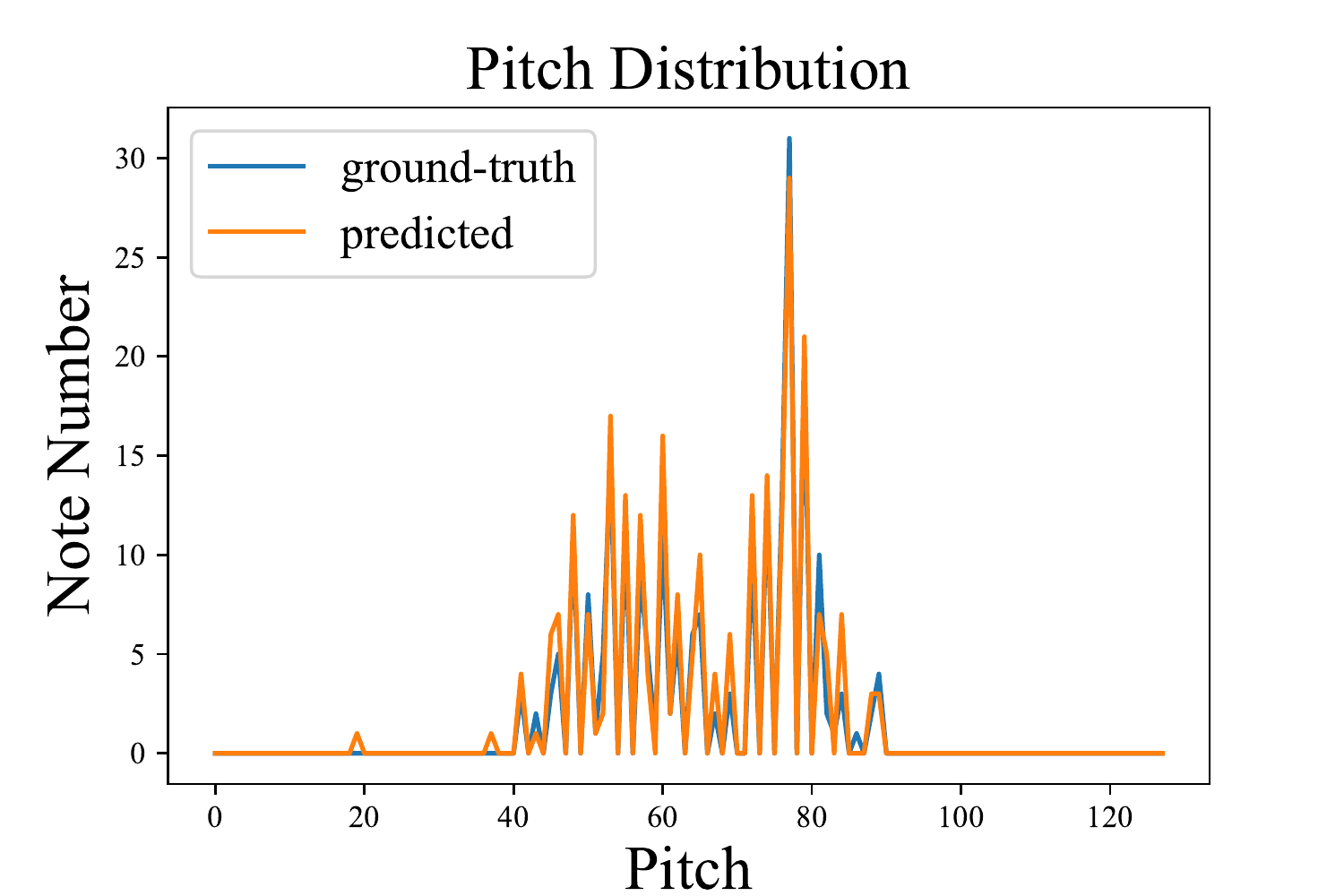}
	    \subcaption{Direct Embedding.}
	\end{subfigure}
	\begin{subfigure}[t]{0.32\textwidth}
	    \captionsetup{justification=centering}
		\centering
	    \includegraphics[width=\textwidth,trim={0cm 0cm 0cm 0cm}, clip=true]{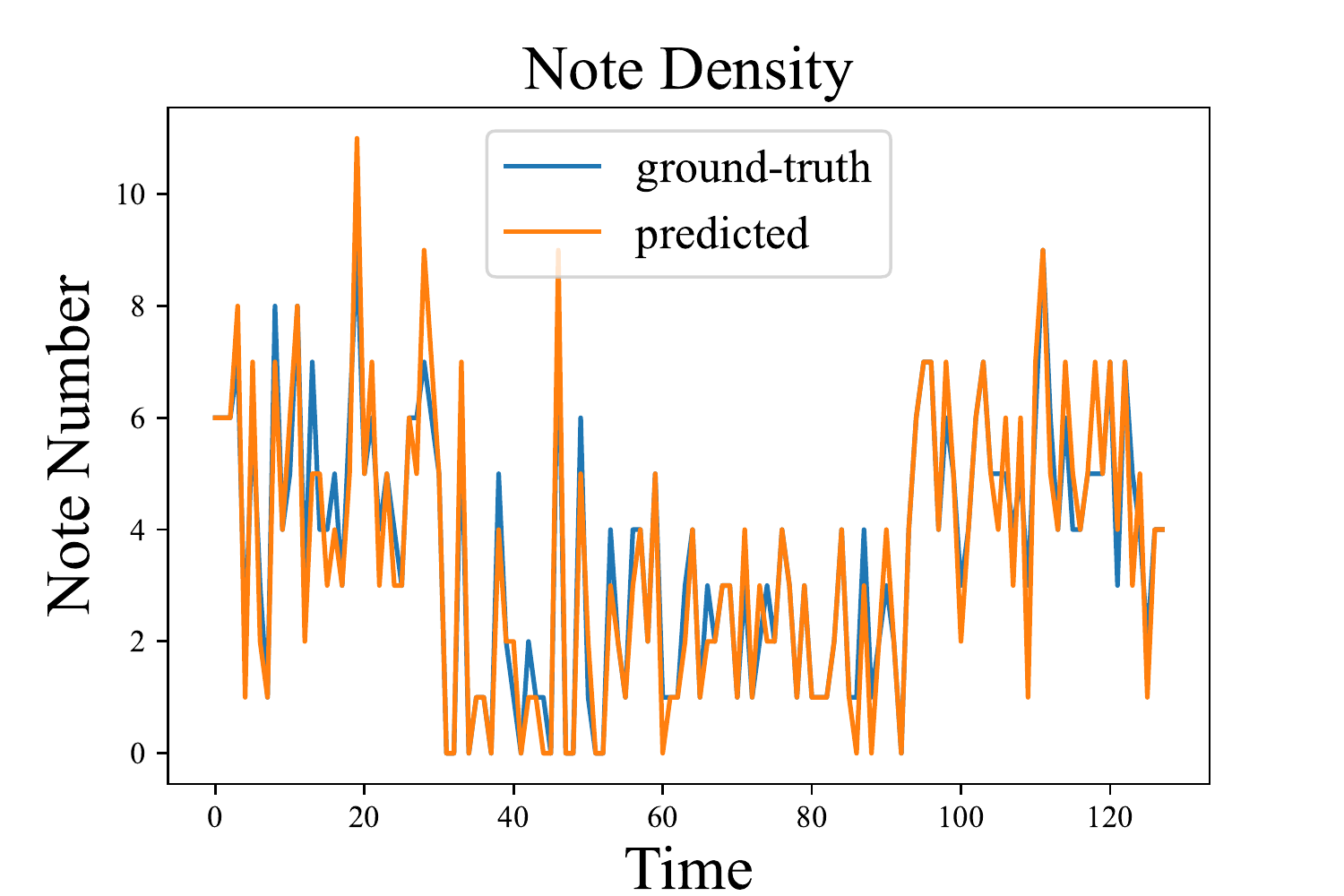}
	    \includegraphics[width=\textwidth,trim={0cm 0cm 0cm 0cm}, clip=true]{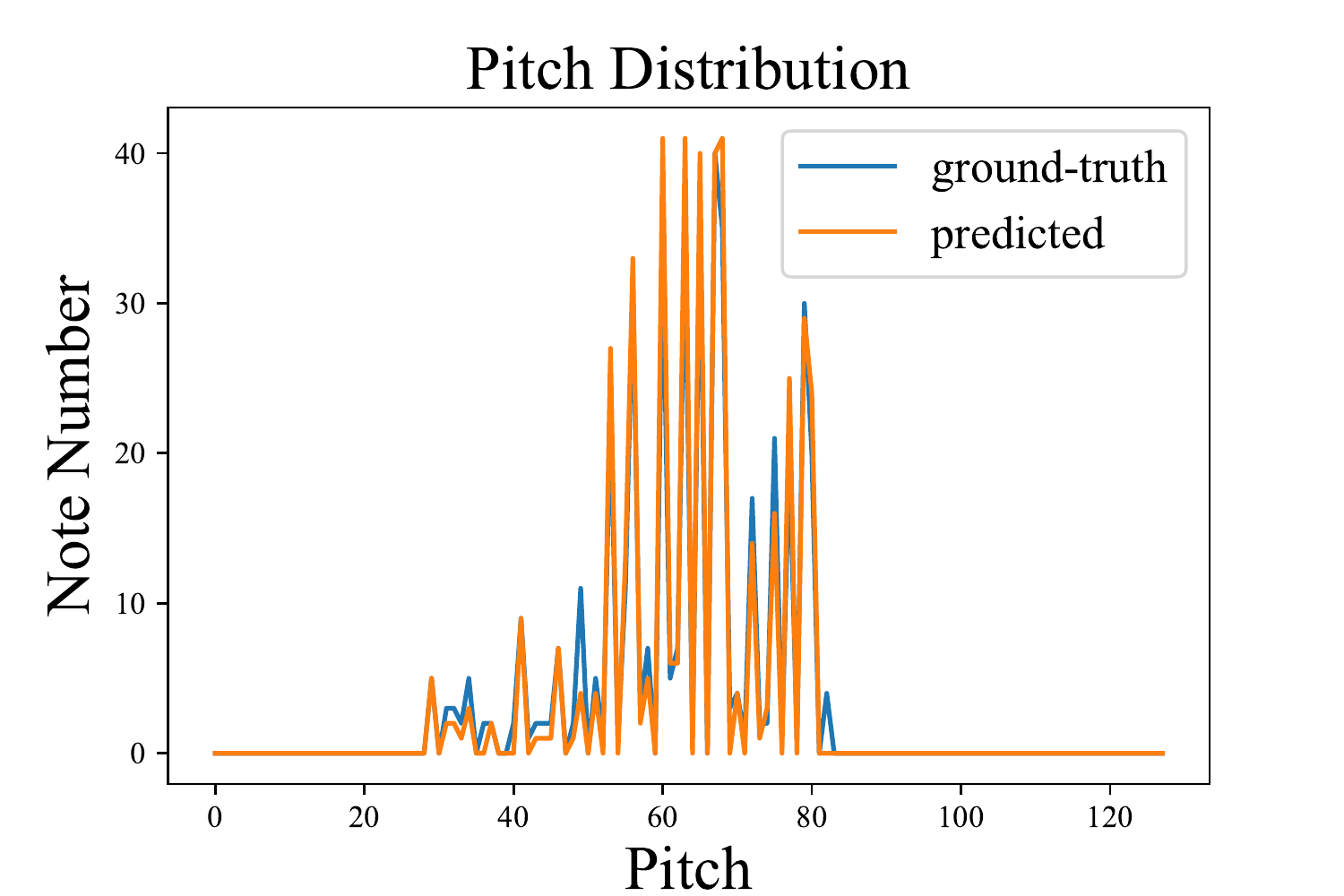}
	    \subcaption{Word Embedding.}
	    \label{fig:control_signals_word_emb}
	\end{subfigure}
    \caption{The visualization of the difference in $\bm{c}_n$ and $\bm{c}_p$ between input and output with different control signal's embedding ways.}
    \label{fig:embedding_ways}
\end{figure}
When involving control signals $\bm{c}_n$ and $\bm{c}_p$ in the diffusion probabilistic model of the pianoroll generation stage, there are several embedding ways: 1) positional embedding: convert the control signals into sinusoidal positional embeddings; 2) direct embedding: regard control signals as vectors and use them as embeddings directly; 3) word embedding: convert the control signals to randomly initialized word embeddings.
As shown in Figure~\ref{fig:embedding_ways}, the ways of direct embedding and word embedding show good performance in terms of the control signals' faithfulness.

\end{document}